\documentclass[twocolumn]{aastex62}

\usepackage{newtxtext,newtxmath}
\usepackage[T1]{fontenc}
\usepackage{ae,aecompl}
\usepackage{graphicx}	
\usepackage{amsmath}	
\usepackage{amssymb}	

\hypersetup{
  colorlinks=true,
  urlcolor=blue,
  citecolor= blue,
  linkcolor = blue
}

\graphicspath{{./}{figures/}}

\accepted{10/02/2019}

\submitjournal{ApJ}

\shorttitle{Dwarfs or giants? }
\shortauthors{Thomas et al.}

\begin{document}

\title{Dwarfs or giants? Stellar metallicities and distances in the Canada-France-Imaging-Survey from $ugrizG$ multi-band photometry}

\correspondingauthor{Guillaume F. Thomas}
\email{guillaume.thomas@nrc-cnrc.gc.ca}

\author{Guillaume F. Thomas}
\affiliation{NRC Herzberg Astronomy and Astrophysics, 5071 West Saanich Road, Victoria, BC, V9E 2E7, Canada}

\author{Nicholaas Annau}
\affiliation{
Department of Physics and Astronomy, University of Victoria, Victoria, BC, V8P 1A1, Canada}

\author{Alan McConnachie}
\affiliation{NRC Herzberg Astronomy and Astrophysics, 5071 West Saanich Road, Victoria, BC, V9E 2E7, Canada}

\author{Sebastien Fabbro}
\affiliation{NRC Herzberg Astronomy and Astrophysics, 5071 West Saanich Road, Victoria, BC, V9E 2E7, Canada}

\author{Hossen Teimoorinia}
\affiliation{NRC Herzberg Astronomy and Astrophysics, 5071 West Saanich Road, Victoria, BC, V9E 2E7, Canada}

\author{Patrick C\^ot\'e}
\affiliation{NRC Herzberg Astronomy and Astrophysics, 5071 West Saanich Road, Victoria, BC, V9E 2E7, Canada}

\author{Jean-Charles Cuillandre}
\affiliation{AIM, CEA, CNRS, Universit\'e Paris-Saclay, Universit\'e Paris Diderot, Sorbonne Paris Cit\'e, Observatoire de Paris, PSL University, F-91191 Gif-sur-Yvette, France}

\author{Stephen Gwyn}
\affiliation{NRC Herzberg Astronomy and Astrophysics, 5071 West Saanich Road, Victoria, BC, V9E 2E7, Canada}

\author{Rodrigo A. Ibata}
\affiliation{Universit\'e de Strasbourg, CNRS, Observatoire astronomique de Strasbourg, UMR 7550, F-67000 Strasbourg, France}

\author{Else Starkenburg}
\affiliation{Leibniz Institute for Astrophysics Potsdam (AIP), An der Sternwarte 16, D-14482 Potsdam, Germany}

\author{Raymond Carlberg}
\affiliation{Departement of Astronomy and Astrophysics, University of Toronto, Toronto, ON M5S 3H4, Canada}

\author{Benoit Famaey}
\affiliation{Universit\'e de Strasbourg, CNRS, Observatoire astronomique de Strasbourg, UMR 7550, F-67000 Strasbourg, France}

\author{Nicholas Fantin}
\affiliation{Department of Physics and Astronomy, University of Victoria, Victoria, BC, V8P 1A1, Canada}

\author{Laura Ferrarese}
\affiliation{NRC Herzberg Astronomy and Astrophysics, 5071 West Saanich Road, Victoria, BC, V9E 2E7, Canada}

\author{Vincent H\'enault-Brunnet}
\affiliation{NRC Herzberg Astronomy and Astrophysics, 5071 West Saanich Road, Victoria, BC, V9E 2E7, Canada}

\author{Jaclyn Jensen}
\affiliation{Department of Physics and Astronomy, University of Victoria, Victoria, BC, V8P 1A1, Canada}

\author{Ariane Lan\c con}
\affiliation{Universit\'e de Strasbourg, CNRS, Observatoire astronomique de Strasbourg, UMR 7550, F-67000 Strasbourg, France}

\author{Geraint F. Lewis}
\affiliation{Sydney Institute for Astronomy, School of Physics, A28, The University of Sydney, NSW 2006, Australia}

\author{Nicolas F. Martin}
\affiliation{Universit\'e de Strasbourg, CNRS, Observatoire astronomique de Strasbourg, UMR 7550, F-67000 Strasbourg, France}

\author{Julio F. Navarro}
\affiliation{
Department of Physics and Astronomy, University of Victoria, Victoria, BC, V8P 1A1, Canada}

\author{C\'eline Reyl\'e}
\affiliation{Institut UTINAM, CNRS UMR6213, Univ. Bourgogne Franche-Comt\'e, OSU THETA Franche-Comt\'e-Bourgogne, Observatoire de Besan\c con, BP 1615, 25010 Besan\c con Cedex, France}

\author{Rub\'en S\'anchez-Janssen}
\affiliation{UK Astronomy Technology Centre, Royal Observatory, Blackford Hill, Edinburgh, EH9 3HJ, UK}

\begin{abstract}
We present a new fully data-driven algorithm that uses photometric data from the Canada-France-Imaging-Survey (CFIS; $u$), Pan-STARRS 1 (PS1; $griz$), and Gaia ($G$) to discriminate between dwarf and giant stars and to estimate their distances and metallicities. The algorithm is trained and tested using the SDSS/SEGUE spectroscopic dataset and Gaia photometric/astrometric dataset. At [Fe/H]$<-1.2$, the algorithm succeeds in identifying more than 70\% of the giants in the training/test set, with a dwarf contamination fraction below 30\% (with respect to the SDSS/SEGUE dataset). The photometric metallicity estimates have uncertainties better than 0.2 dex when compared with the spectroscopic measurements. The distances estimated by the algorithm are valid out to a distance of at least $\sim 80$ kpc without requiring any prior on the stellar distribution, and have fully independent uncertainities that take into account both random and systematic errors. These advances allow us to estimate these stellar parameters for approximately 12 million stars in the photometric dataset. This will enable studies involving the chemical mapping of the distant outer disc and the stellar halo, including their kinematics using the Gaia proper motions. This type of algorithm can be applied in the Southern hemisphere to the first release of LSST data, thus providing an almost complete view of the external components of our Galaxy out to at least $\sim 80$ kpc. Critical to the success of these efforts will be ensuring well-defined spectroscopic training sets that sample a broad range of stellar parameters with minimal biases. A catalogue containing the training/test set and all relevant parameters within the public footprint of CFIS is available online.
\end{abstract}


\section{Introduction}

 \begin{figure*}
\centering
  \includegraphics[angle=0, clip, width=18cm]{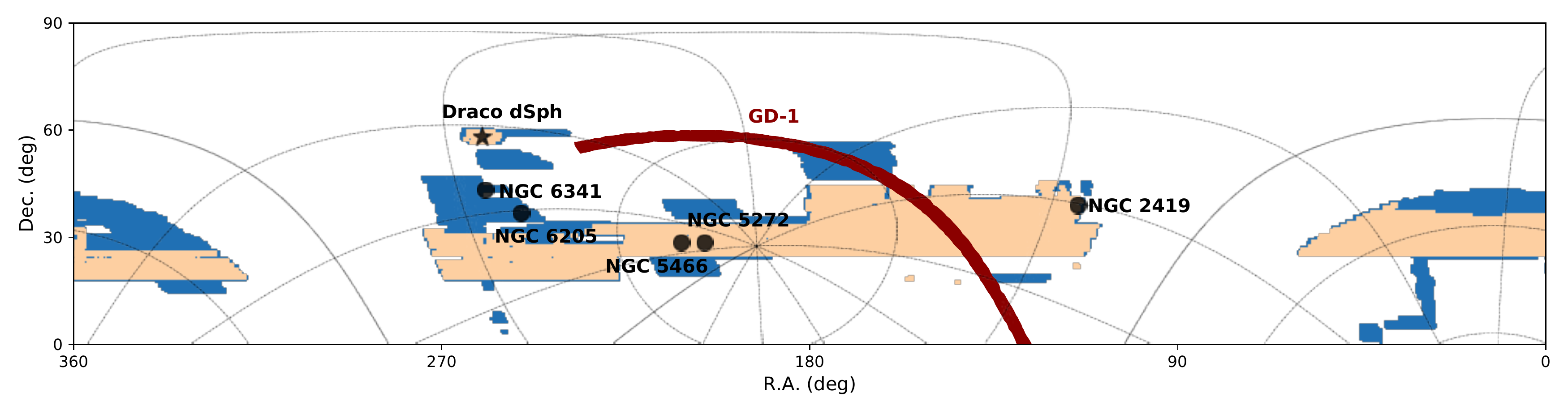}
   \caption{Current footprint of the CFIS-u survey (blue areas) on an equatorial projection. The $\sim 2,600$ deg$^2$ of the public area of CFIS-u is shown by the light orange area. The grey lines show the Galactic coordinates with the solid lines showing the Galactic Plane and the Galactic minor axis. The different satellites within the footprint are used to validate the distance estimated by our algorithm on Section \ref{sec_GC} are also indicated in the figure.}
\label{footprint}
\end{figure*}

The advent of the European Space Agency's Gaia satellite has yielded accurate proper motion for stars brighter than $G\sim 21$ \citep{gaiacollaboration_2018c}. However, to transform this angular velocity into a tangential velocity, accurate distances are required. With the parallaxes measured from Gaia, the distances of the stars in the Solar vicinity can be measured with high precision out to a few kiloparsecs ($sim$ 10\% at 1.5 kpc). Despite this very impressive number, the distances of the large majority of the 1.3 billion stars present in the Gaia catalog cannot be accurately inferred using only Gaia parallaxes. For example, a main sequence star at 3 kpc has a parallax uncertainty of $\simeq 10\%$, and at 7 kpc the uncertainty on the parallax is of the same order as the parallax measurement itself \citep{bailer-jones_2013,ibata_2017b}. Therefore, distances to stars in the outer disc of the Galaxy and in the stellar halo cannot be accurately measured by direct inversion of the parallaxes \citep{bailer-jones_2015,luri_2018a}.

Several methods have been developed to infer statistically the distances of these stars using assumptions made on the global distribution of the stars in the Galaxy \citep[e.g.][]{bailer-jones_2015,bailer-jones_2018,queiroz_2018,anders_2019,pieres_2019}. However, the actual distribution of stars in the Galaxy, especially in the stellar halo, is still not known precisely, and different tracers yield different distributions \citep{thomas_2018a,fukushima_2019}. Therefore, the correct prior to adopt on the ``expected'' distribution of the stars is not obvious. Moreover, the spatial distribution of stars found using distances estimated by these methods depends sensitively of the adopted prior \citep{hogg_2018}.

To overcome this problem, spectrophotometric methods have been developed to infer stellar distances \citep[e.g.][]{xue_2014,coronado_2018,mcmillan_2018,queiroz_2018,hogg_2018}. However, these methods require expensive spectroscopic observations. Moreover, the current generation of spectroscopic surveys do not exploit the full depth of Gaia. \citet{juric_2008} and \citet{ivezic_2008} developed a method to estimate the distance and metallicity of stars using the SDSS $u$, $g$, $r$, and $i$ bands that circumvents the need for spectroscopy. This method was revisited by \cite{ibata_2017b}. Inherent to this method is the assumption that all stars are main sequence stars. Thus a giant with the same color as a main sequence star will be estimated to be much closer than its true distance by several orders of magnitude.

To study in detail the chemical distribution and kinematics of the outer disk, the complex structure of the stellar halo, and the interface region between the disk and the stellar halo, it is crucial to measure the distance of the stars, including the giants, over the full depth of Gaia.

In this paper, we present a new technique to estimate distances and metallicities, that is based heavily on the methods of \citet{juric_2008, ivezic_2008} and \citet{ibata_2017b}, and which incorporates Machine Learning techniques. This fully data-driven algorithm first discriminates between dwarfs and giants based on photometry alone, and then estimates the distances and metallicities for each set using the same photometry. More specifically, we use multi-band photometry provided by the Canada-France-Imaging Survey (CFIS), Pan-STARRS 1 (PS1) and Gaia. This dataset is presented in Section \ref{data}. The architecture of the algorithm and its calibration are detailed in Section \ref{algorithm}. The accuracy and the biases of the algorithm are tested using independent datasets in Section \ref{verif}. We apply the algorithm to 12.8 million stars in the CFIS footprint in Section \ref{map_metal} and use these data to map the mean metallicity of the Galaxy. Finally, in Section \ref{conclusion}, we discuss the applicability of this type of algorithm to future datasets, including LSST, and discuss the scientific opportunities it presents.

\section{Data} \label{data}

 \begin{figure}
\centering
  \includegraphics[angle=0, clip, width=7.5cm]{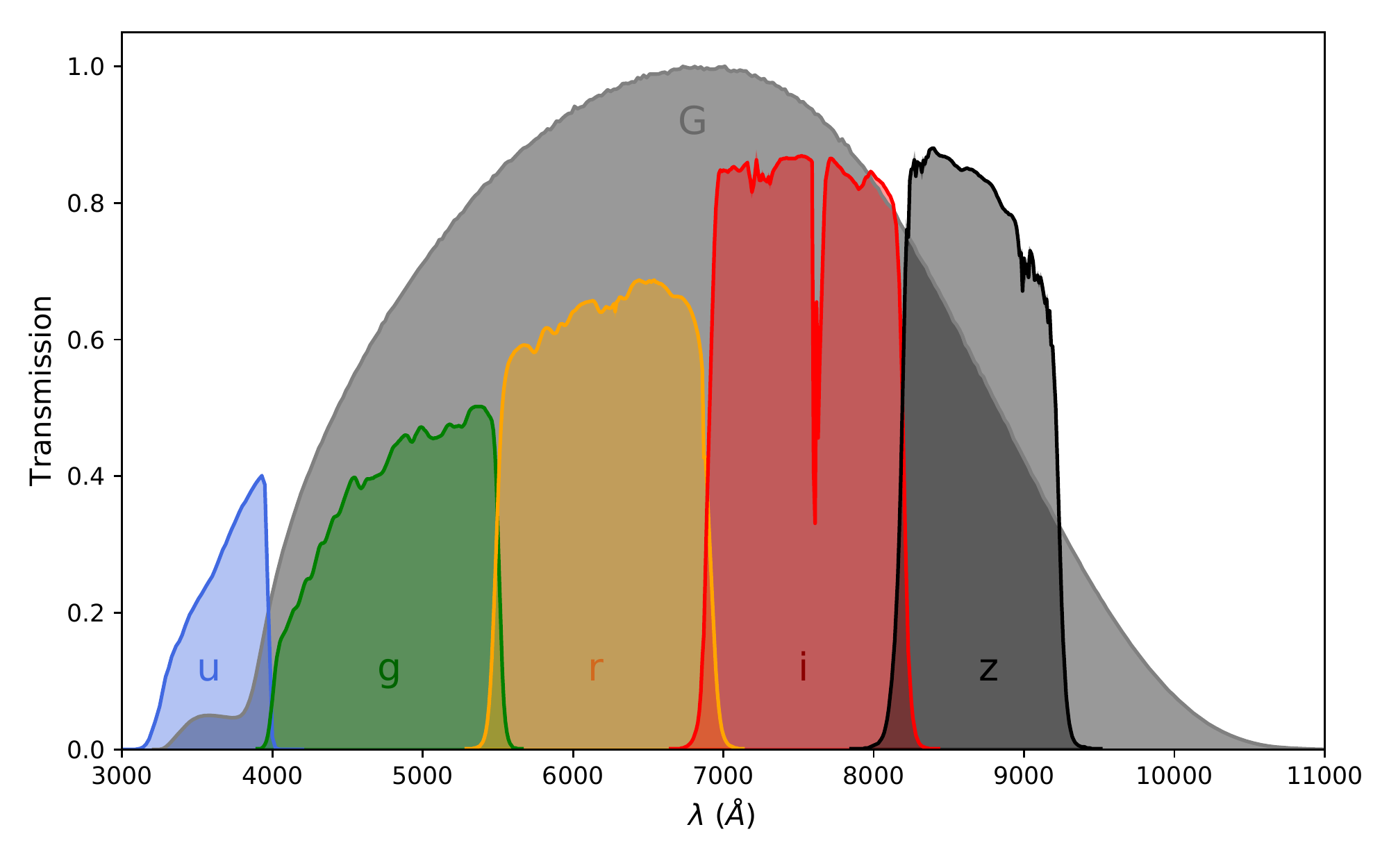}
   \caption{The relative transmission of the photometric passbands used in this analysis.}
\label{transmission}
\end{figure}

 \begin{figure*}
\centering
  \includegraphics[angle=0, clip, viewport= 0 0 800 755 ,width=14.5cm]{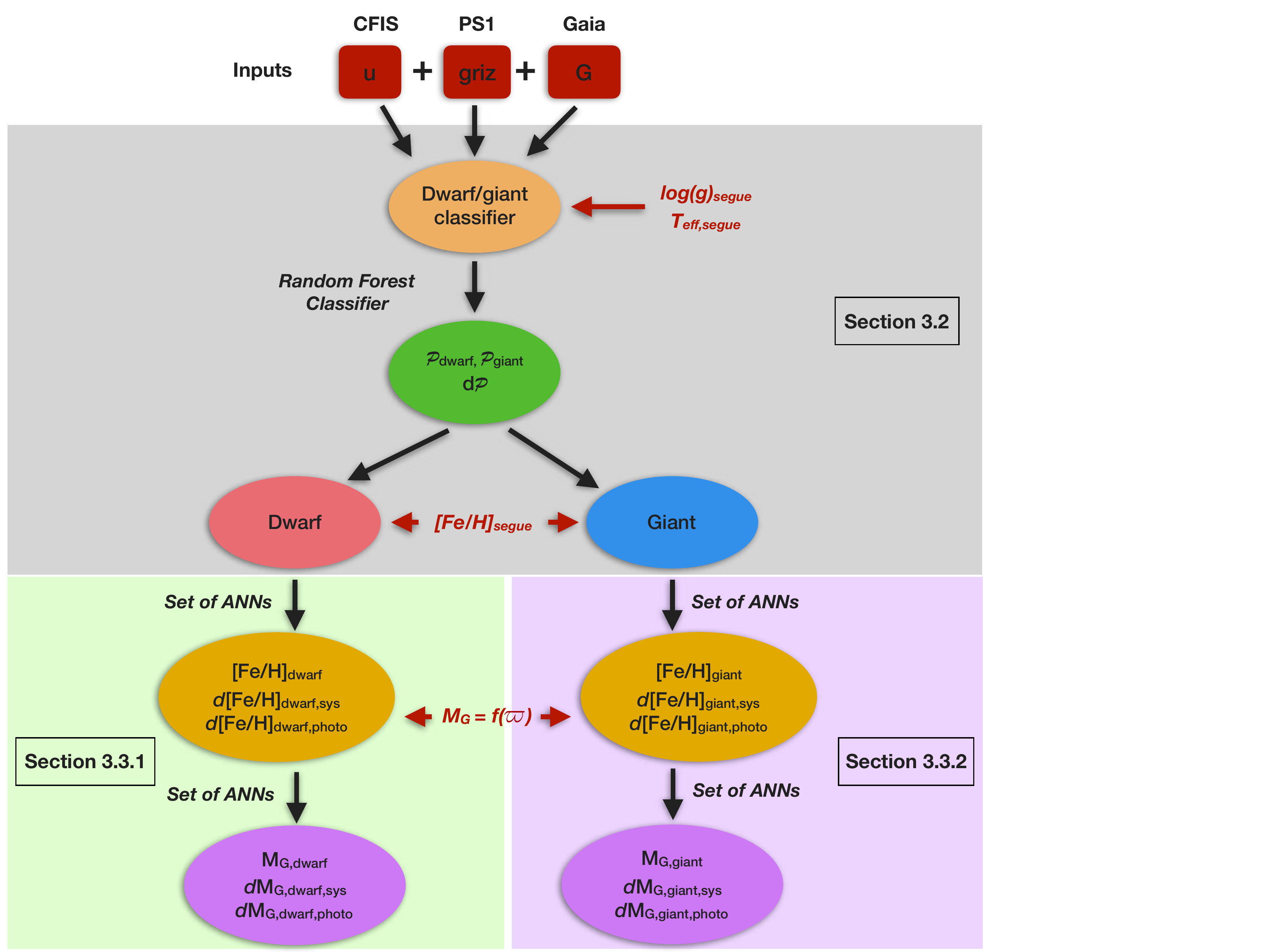}
   \caption{Schema of the algorithm described in Section \ref{algorithm}. The input parameters are the photometric measurements from various surveys described in Section \ref{data}. Parameters in red are used for training. Dwarf - giant classification is done using  a {\it Random Forest Classifier}. For each class, a set of {\it Artificial Neural Networks} (ANNs) is used to compute the photometric metallicity. Then, another set of ANNs is used to compute the absolute magnitude in the $G$-band, which provides the distances to the stars.}
\label{schema}
\end{figure*}

The photometric catalog used in this study  (hereafter referred to as the {\it main} catalog) is a merger of the {\it u}-band photometry of CFIS \citep{ibata_2017a}, with the {\it griz}-bands from the {\it mean PSF} catalog of the first data release of PS1 \citep{chambers_2016} (PS1), and the {\it G}-band from the Gaia DR2 catalog \citep{gaiacollaboration_2018}\footnote{The $y$-band  from PS1 and the  $G_{BP}$ and $G_{RP}$ bands of Gaia are not used in this study because the large photometric uncertainties for stars fainter than $G>19.5$ lead to large uncertainties in the derived photometric metallicities and distances.}. Spatially, the survey area is limited by the current coverage of CFIS-$u$ ($\sim 4000$ deg$^2$ of the northern hemisphere, eventually covering 10,000 deg$^2$ at the end of the survey), since Gaia is all-sky and PS1 covers the entire sky visible from Hawaii. Photometrically, the depth is limited by the Gaia $G$-band. The total and public footprint of CFIS-u at the time of this analysis are shown in Figure \ref{footprint} in blue and orange, respectively.

The normalized transmission of the different filters that constitute the {\it main} catalog are shown in Figure \ref{transmission}. The different filters cover a range of wavelength from the near-UV to the near-IR (specifically, from $\lambda \simeq 3200\, \mathrm{\mathring{A}}$ to $11000\, \mathrm{\mathring{A}}$). This large photometric baseline is useful to provide information on the overall shape of the spectral energy distribution (SED) of the stars, and we note that the overlap between different filters in some spectral regions, e.g. around $ 5500\, \mathring{A}$, provides extremely valuable information in a short range of wavelength, comparable to extremely low-resolution spectroscopy. Indeed, it is expected that the absorption lines present in these overlap regions, such as the FeH-I and Ca-I, around $ 5500\, \mathring{A}$, will have a stronger impact on the algorithm than other absorption lines located in the middle of a filter for which their signal is harder to disentangle from the rest of the SED. All of these features are put together by the algorithm described in the next section to obtain dwarf/giant classification, the metallicity and the distance (absolute magnitude) of the stars. 

For star-galaxy classification, we adopt the PS1 criteria, $r_{PSF}-r_{Kron}<0.05$. As pointed out by as \citet{farrow_2014}, star-galaxy separation using this criterion become unreliable for stars fainter than $r_{PSF} = 21$. Since our catalog is effectively limited by the Gaia G-band limiting magnitude ($G \simeq  20.7$), the majority of our sources (more than 99.9 percent) have $r_{PSF} < 21$. Therefore, star-galaxy misclassification has negligible impact on the results of this study. 

We use extinction values, $E(B-V)$, as given by \citet{schlegel_1998}. The CFIS footprint is at relatively high Galactic latitude, $| b |>19 \degr$, and most stars are moderately distant, so we can reasonably assume that all of the extinction measured in the direction of a star is in the foreground of the star. While this assumption is clearly hazardous for the closest stars, we will see below that it does not have a large impact on our results when we trace the chemical distribution of the disc of the Milky Way \citep[see also][]{ibata_2017b}. We adopt the reddening conversion coefficients for the {\it griz}-band of PS1 given by \citet{schlafly_2011} for a redenning parameter of $R_\mathrm{v}=3.1$. As in \citet{thomas_2018a}, we assume that the conversion coefficient of the $u$-band of CFIS is similar to the coefficient of the SDSS $u$-band from \citet{schlafly_2011}. For the Gaia $G$ filter, we follow \citet{sestito_2019} by adopting the coefficient from \citet{marigo_2008} (based on \citet{evans_2018}).


\section{Method} \label{algorithm}

\subsection{Overview}

Figure \ref{schema} provides a schematic overview of the algorithm that we have developed. This algorithm only uses the photometric data described in the previous section to first disentangle giants from dwarfs using a {\it Random Forest Classifier} (RFC) \citep{breiman_2001}, as detailed in Section \ref{sec_DG}. Once this classification is done, two sets of {\it Artificial Neural Networks} (ANNs), one for the stars identified as dwarfs (Section \ref{sec_dwarf}) and another for the stars identified as giants (Section \ref{sec_giant}), are used to determine the photometric metallicity. Finally, another two  sets of ANNs are used to determine the absolute magnitude of the two groups of stars in the $G$-band.

To calculate the uncertainties on the different parameters estimated by our algorithm ($P_{Dwarf}$, $P_{Giant}$, [Fe/H] and $M_G$) caused by the photometric uncertainties of the different bands, we generate 20 Monte-Carlo realizations for each star. We also conducted 100 and 1,000 Monte-Carlo realizations for a sub-sample of 50,000 randomly selected stars. For these, we obtained final uncertainties that were typically $< 1\%$ different than what we obtained using only 20 realizations. As such, we proceeded with 20 Monte-Carlo realizations for the entire sample in order to save expensive computational time. For each realization, we select a magnitude in each band from a Gaussian distribution centered on the quoted magnitude and with a standard deviation equal to the uncertainty on the magnitude. The uncertainties on the derived parameters are set equal to the standard deviation for each parameter from these 20 realizations.

The CFIS-$u$ footprint, which defines the spatial coverage of this study, includes a large number of stars observed by SDSS/SEGUE \citep{yanny_2009} and for which spectroscopic data are available. We use those stars with good quality spectroscopic measurements as training sets for the first two components of our algorithm. It is advantageous to use as large a training set as possible; therefore, we select $74,442$ SDSS/SEGUE stars present in the CFIS-u footprint that have a spectroscopic signal-to-noise ratio of $SNR \geq 25$. This threshold was chosen because at lower SNR, the distribution of the uncertainties on the parameters given by the SEGUE  Stellar Parameters  Pipeline (SSPP) as a function of the SNR is irregular, indicating that the parameters are poorly defined. Moreover, more than 96\% of the stars with $SNR<25$ have a parallax measurement with poor precision ($>20\%$) and these would not be used to calibrate the photometric distance relation even if they passed the SNR cut. For the third component of our algorithm (determination of the absolute magnitude), we use parallax information from Gaia (discussed later).

 \begin{figure}
\centering
  \includegraphics[angle=0, clip , width=8.5cm]{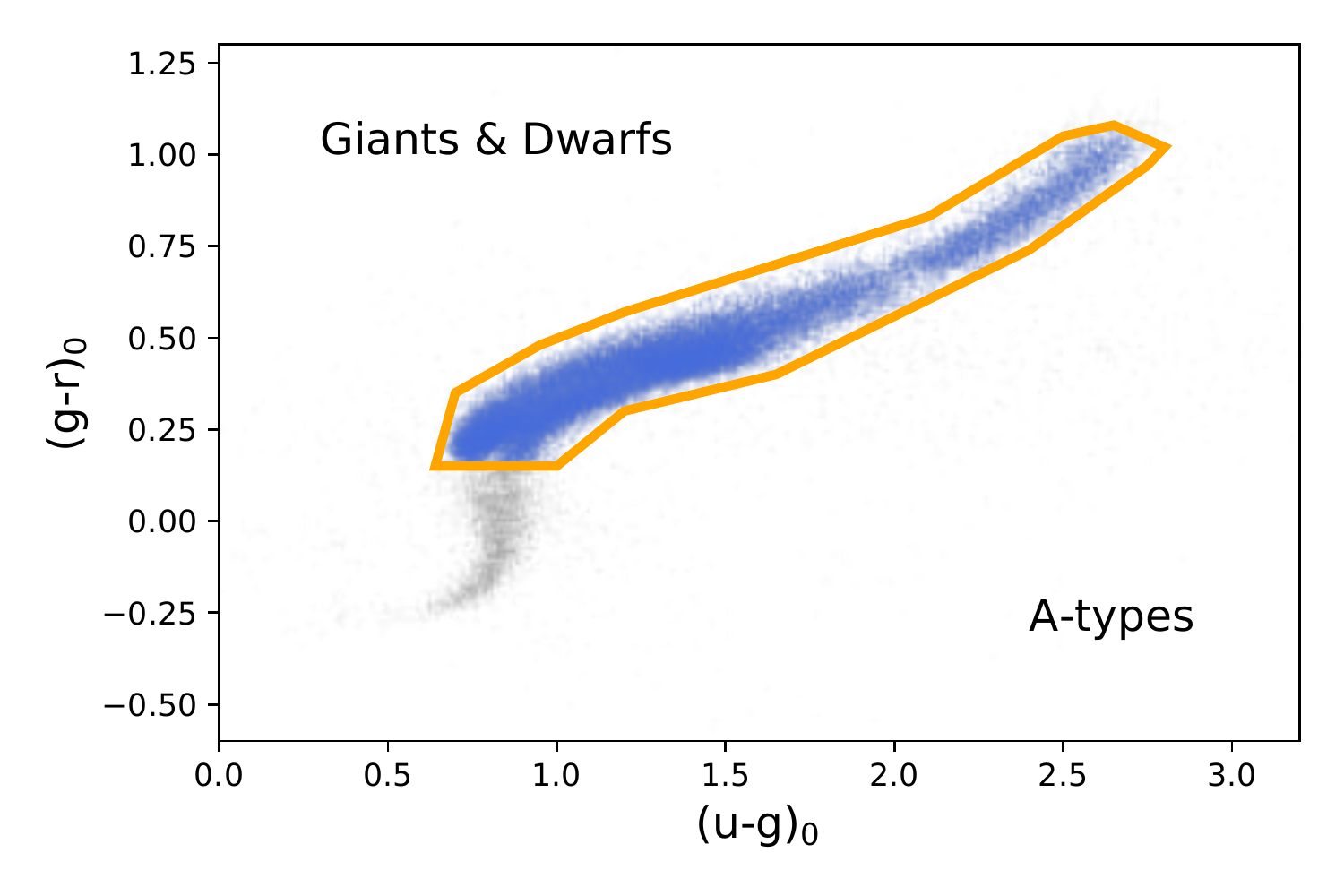}
   \caption{Color-color diagram of the SDSS/SEGUE stars that have $SNR \geq 25$. The orange polygon corresponds to the locus of main-sequence and RGB stars in this color-color plane. This selection box removes A-type stars and white dwarfs from the subsequent analysis.}
\label{col-col}
\end{figure}
 
We first perform a color-color cut to remove A-type stars (which lie in the ``comma-shaped'' region in the color-color diagram of Figure \ref{col-col}) and white dwarfs, that is defined by inspection of the Figure \ref{col-col} and shown as the orange box. The presence of the Balmer jump in the $u$ band for these stars means that they have a more complex photometric behavior compared to the other stars present in this color-color diagram, and the algorithms are significantly simplified if we remove them from consideration. We note that A-types stars in CFIS have been studied extensively in previous works \citep{thomas_2018a,thomas_2019}, and an analysis of the white dwarfs is in preparation (Fantin et al. in prep.). Imposing this color-color selection\footnote{The  $(u_0-g_0, g_0-r_0)$ vertices of this selection are : (0.64, 0.15), (1.0 0.15), (1.2, 0.3), (1.65, 0.4), (2.4, 0.74), (2.75, 0.97), (2.8, 1.02), (2.65, 1.08), (2.5, 1.05), (2.1, 0.83), (1.2 0.57), (0.95, 0.48), (0.7, 0.35).} on the SDSS/SEGUE spectra lead to a catalog of $\sim 42,800$ stars for which we have astrometric, photometric and spectroscopic information. This color-color cut represent the first step of our procedure, and is applied to botht the test/training set and the final main catalog (see Section \ref{all_cat}).

\subsection{Dwarf - giant classification} \label{sec_DG}

Here, we describe the method used to disentangle Red Giant Branch stars (RGBs) from main-sequence stars (MS)\footnote{Hereafter, we refer to the RGBs as {\it giants} and to the MSs as {\it dwarfs}.}. To perform this classification we use a Random Forest Classifier, whose inputs are the $(u-g)_0$, $(g-r)_0$, $(r-i)_0$, $(i-z)_0$ and $(u-G)_0$ colors normalized to have a mean of zero and a standard deviation of one and the outputs are the probability of each to be a dwarf, $P_{dwarf}$, or a giant, $P_{giant} \equiv 1-P_{dwarf}$. It is worth noting here that RFC does not necessary request a normalization of the inputs, but it is strongly suggested for an ANN. Therefore, to be consistent with the different steps of the algorithm, we use the normalized inputs for the dwarf/giant classification.

\begin{figure}
\centering
  \includegraphics[angle=0, clip, width=7.5cm]{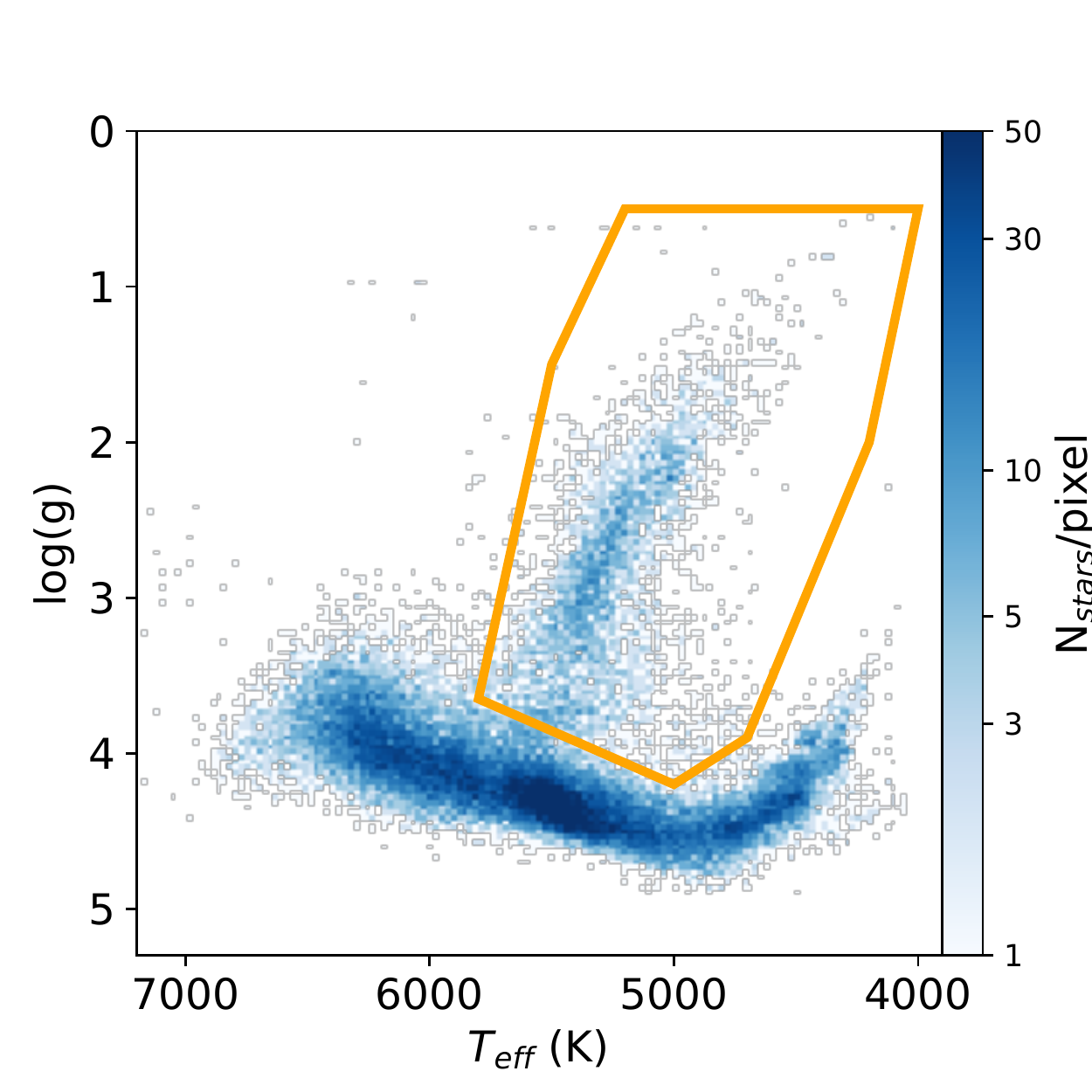}
   \caption{$Kiel$ diagram showing the distribution of stars in the SDSS/SEGUE catalog used to classify dwarfs and giants. The orange polygon shows those stars we ``define'' as giants.}
\label{den_HR}
\end{figure}

According to \citet{lee_2008}, the typical internal uncertainties obtained by the SSPP on the adopted surface gravity (\textsf{loggadop}) is $\sim 0.19$ dex. Therefore, we keep only the SDSS/SEGUE stars that have uncertainties $\delta \log(g) \leq 0.2$. The final SDSS/SEGUE catalog used to train/test the dwarf - giant classifier contains $41,062$ stars. This catalog is shown as a $Kiel$ diagram in Figure \ref{den_HR}, where we define the stars lying in the orange polygon as {\it giants} and all other stars as {\it dwarfs}. Note that we could in principle use Gaia DR2  parallax measurements to classify the stars as dwarfs or giants; however as we will show later, the parallax measurements of Gaia DR2 are not precise enough, especially for stars with $[$Fe$/$H$]<-1.0$, many of which are generally located at large distances and have poor parallax precision.

 \begin{figure*}
\centering
  \includegraphics[angle=0, clip, viewport= 0 0 1020 580 , width=15cm]{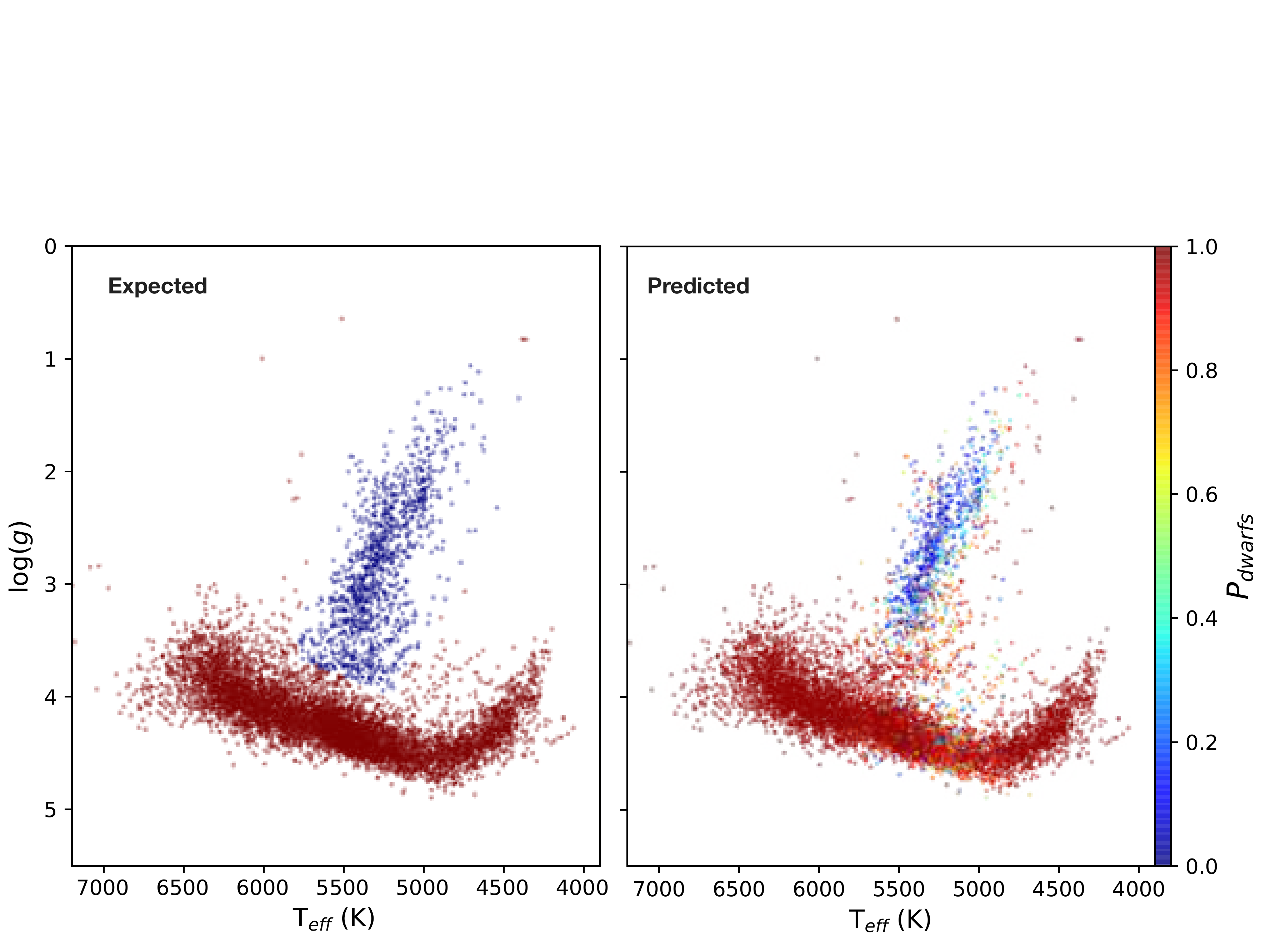}
   \caption{$Kiel$ diagrams where each point correspond to a star in the test set color-coded by the probability of it being a dwarf. The expected probability, or the actual class of stars, is show in the left panel and the predicted probability from our algorithm is shown in the right panel.}
\label{HR}
\end{figure*}

We note that our ``giant'' selection contains a large majority of RGB stars but also sub-giants stars, with an effective temperature between $5000\leq $ T$_{eff}$ (K) $\leq 6000$ and  surface gravity between $3.2 \leq \log(g) \leq 3.9$. Even on a $Kiel$ diagram, it is hard to define a strict limit between dwarfs, sub-giants and giants, especially when we include the uncertainties on the surface gravity which acts to blur any boundaries we adopt. The addition of a sub-giant class adds to the complexity of the algorithm and does not resolve the underlying problem of the ``fuzzyness'' between classes. For this reason, we do not consider a specific class of sub-giant stars. We also note that our definition of giants does not include the majority of Asymptotic Giant Branch stars (AGBs). However, there are very few AGBs in the SDSS/SEGUE catalog, and so we are unable to train the algorithm to identify them. Like all supervised machine learning algorithms, we are ultimately limited by the representative nature of our training set. AGB stars are, however, very rare, and so their absence from our training set does not have a major statistical effect on our results.

We create a training and a test set from the SDSS/SEGUE catalog, composed of a randomly selected  80\% and 20\% of the sample, respectively. The training set is used to find the best architecture by a {\it k-fold} cross validation method with five sub-sample. It is then used to find the best parameters of the RFC that are then applied to the test set to check that the statistics of the two samples are similar. This technique prevents over-fitting to the training set. We use the \textsf{sklearn} python package to find the best parameter of the RFC.

 \begin{table}
 \centering
  \caption{Completeness and purity of the dwarf and giant classes for the test sample. The results are statisitically the same for the training set, demonstrating that we are not over-fitting to the training set. The first column refers to the true fraction of stars actually classified as dwarfs or giants in the test set.}
  \label{table_class}
  \begin{tabular}{@{}lccc@{}}
  \hline
   Class & Fraction & Completeness & Purity  \\
    \hline
   Dwarf & 0.86 & 0.96 & 0.93  \\
   Giant & 0.14 & 0.57 & 0.70 \\
\hline
\end{tabular}
\end{table}

 \begin{figure*}
\centering
  \includegraphics[angle=0, clip, width=15.0cm]{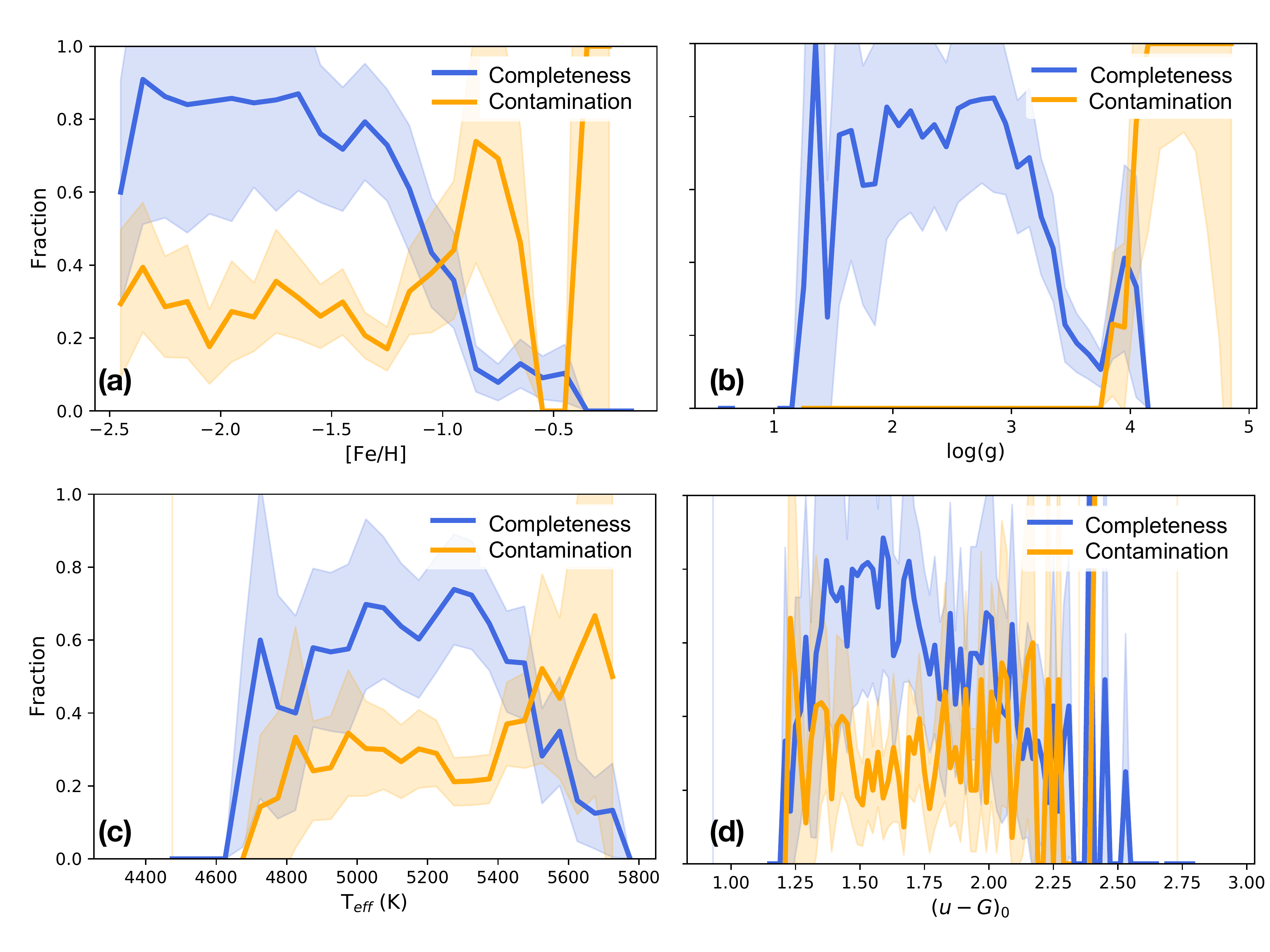}
   \caption{Completeness (blue) and contamination (orange) of stars classified as giants by the $RFC$ in the test set as function of (a) spectroscopic metallicity, (b) surface gravity, (c) effective temperature, (d) color $(u-G)_0$. }
\label{bias}
\end{figure*}

 \begin{figure}
\centering
  \includegraphics[angle=0, clip,  viewport= 0 0 1020 650, width=7.5cm]{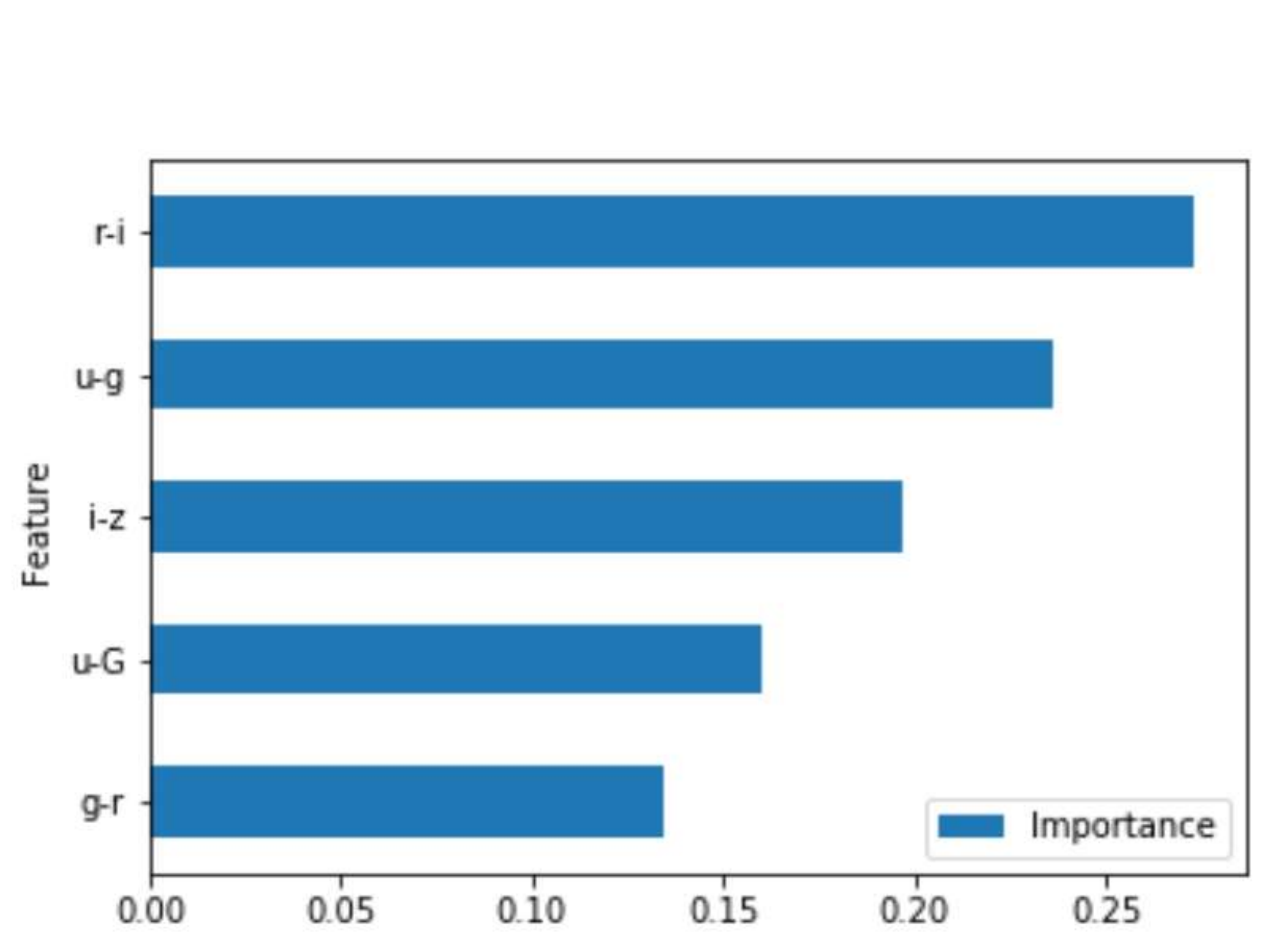}
   \caption{Relative importance of each color in the classification of dwarfs and giants by the RFC.}
\label{feature_imp}
\end{figure}

The completeness and the purity of the dwarf and giant classes (populated by stars with $P_{dwarf}$/$P_{giant}>0.5$) for the test set are shown in Table \ref{table_class}. The values for the two classes are similar to those for the training set, which indicates that there is no over-fitting of the data. The RFC classifies correctly the large majority of the dwarfs ($96\%$), with less than 7\% contamination by giant stars (note, that since dwarfs make up 86\% of our training/test set, then this implies a factor of 2 improvement over random chance). For the giants, slightly more than half are correctly classified, with relatively low contamination, $\sim 30\%$. Thus, our completeness is approximately 4 times better than ``random'', and our contamination is nearly 3 times lower than ``random''. Figure \ref{HR} is a $Kiel$ diagram of the expected/predicted probability for each star to be a dwarf, and we see that most of the contamination of the dwarfs is from sub-giants, which are preferentially classified as dwarfs instead of giants.

It is worth nothing here that, in Figure \ref{HR}, the surface gravity from SDSS/SEGUE decreases with the temperature for the main sequence stars with an effective temperature lower than 4,800 K. This is unexpected when compared to theoretical predictions, and it is likely a consequence of the poor determination of the surface gravity by the SSPP in this region. However this has no impact on our classification, since all the stars in this region are correctly identified as dwarfs by the algorithm.

The detailed performance of our classification scheme as a function of metallicity, surface gravity, effective temperature and $(u-G)_0$ color is shown in Figure \ref{bias}. The completeness and contamination are mostly constant in the range of temperature and color covered by the giants. The completeness and contamination are also constant with metallicity up to [Fe/H]$\simeq-1.2$, after which the completeness drops rapidly (from 70\% at [Fe/H]$=-1.3$ to 20\% at [Fe/H]$=-1.0$), correlated with a dramatic increase in the contamination. We conclude that the classification works well for giants with metallicities below [Fe/H]=$-1.2$, but fails to identify the most metal-rich giants. There is also a drop in the completeness of the giants at a surface gravity of $\log(g)$ of 3.3, which corresponds to the sub-giants being preferentially classified as dwarfs.

The relative importance of each photometric color in the classification scheme, computed using the \textsf{feature importances} method implemented in the \textsf{sklearn} package, is shown in Figure \ref{feature_imp}. This method uses the weight of each feature in each node of the different trees of the RFC to measure the relative importance of such feature for the classification. The most important feature is the $(r-i)_0$ color, with around 1/4 of the information used to classify the stars coming from it. This is not surprising since this color is a good indication of the effective temperature, therefore we assume that this color is being used to select the temperature ranges that preferentially contains giants ($4,900\leq T_{eff} \leq 5,500$ K). The second most important feature is $(u-g)_0$, which has a tight correlation with the metallicity of MSs \citep{ivezic_2008,ibata_2017b}. A similar correlation exists for the giants, albeit one with a different zero point \citep{ibata_2017b}. The other colors, that account for $\sim 50\%$ of the relative importance, presumably give additional minor complementary information (on the metallicity, temperature, surface gravity) to disentangle the dwarfs from the giants, using the full shape of the spectral energy distribution (SED).

 \begin{figure*}
\centering
  \includegraphics[angle=0, clip, width=15.0cm]{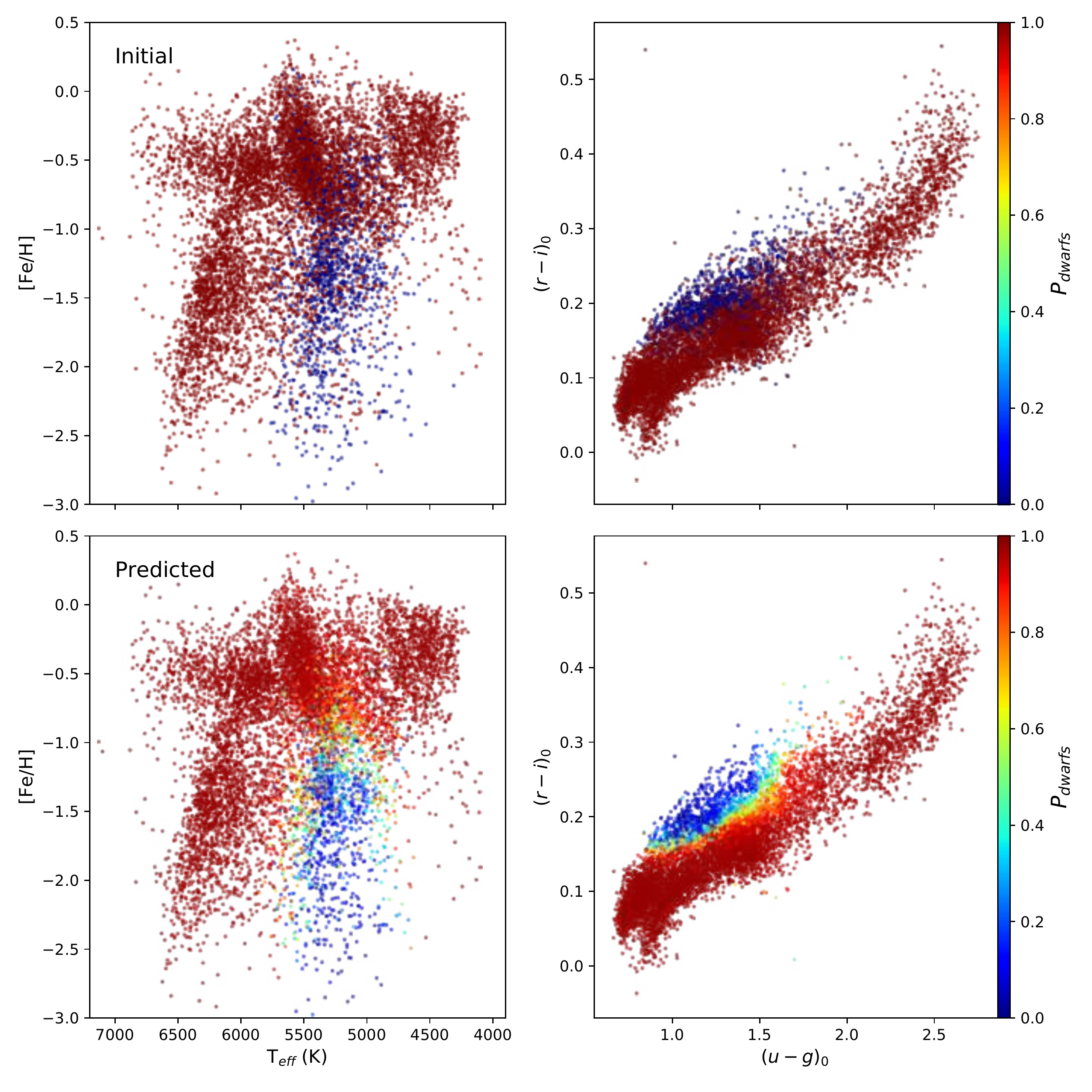}
   \caption{The left panels show effective temperature-metallicity diagrams, and the right panels show a $(u-g)_0 - (r-i)_0$ diagram. The top panels show the distribution of true dwarfs (red) and giants (blue) of the test set. The lower panel show the same distributions where the stars are color-coded as a function of their probability to be dwarfs according to the algorithm.}
\label{teff_feh}
\end{figure*}

 \begin{figure}
\centering
  \includegraphics[angle=0, clip, width=8.5cm]{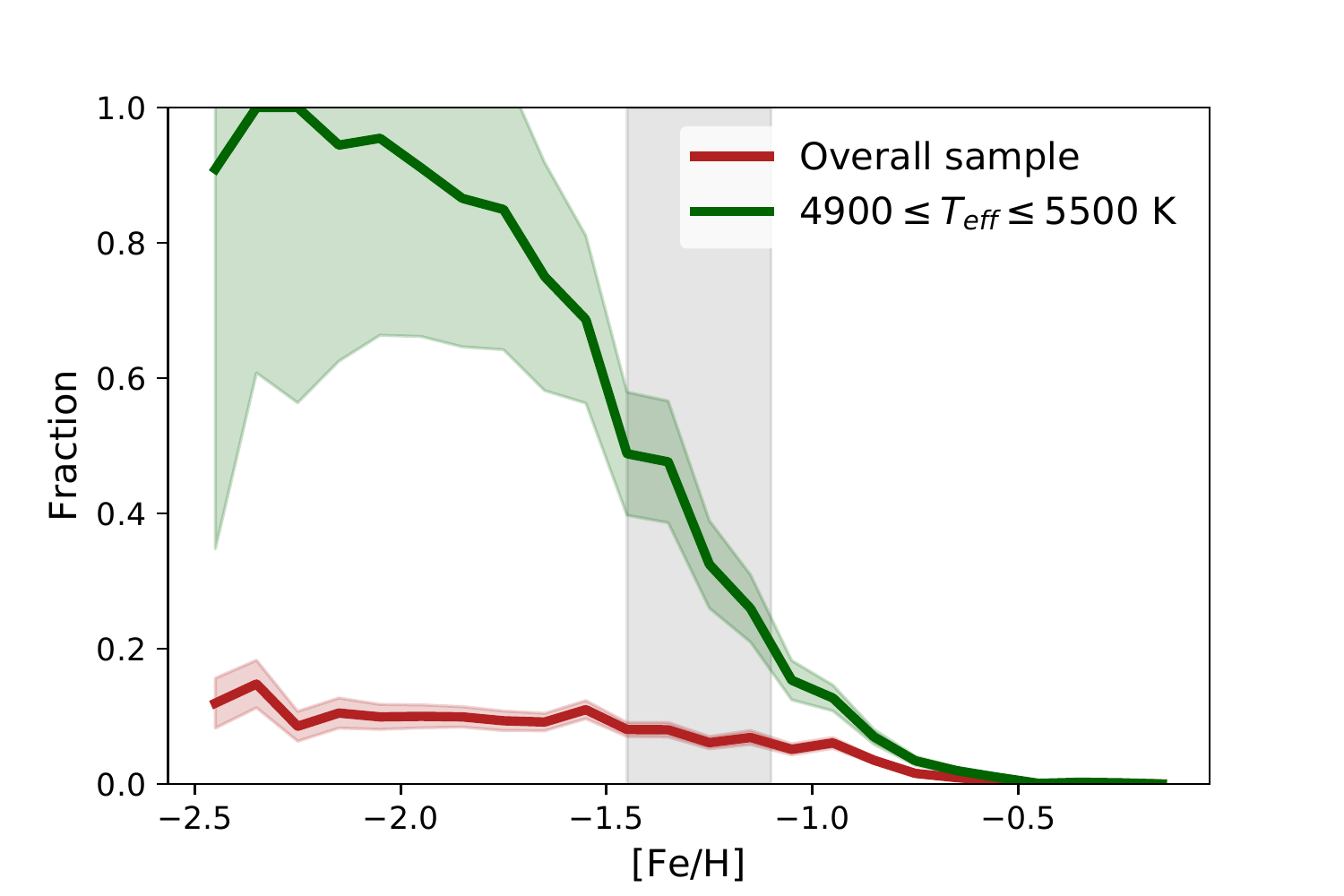}
   \caption{Fraction of actual dwarfs misidentified as giants as a function of metallicity, for the overall test sample (red), and for a narrower range of effective temperature between $4,900\leq $T$_{eff} \leq 5,500$ K (which is the temperature range of the giants). The gray shadowed area highlights the metallicity region where more than 50\% of dwarfs and giants are correctly identified.}
\label{frac_miss}
\end{figure}

From Figure~\ref{feature_imp}, we conclude that the dwarfs/giant classification primarily uses photometric features that trace temperature and metallicity. This becomes more clear in Figure \ref{teff_feh}, where the locus of giants is obvious on the effective temperature-metallicity diagram (left panels), or on a color-color diagram using the two most important photometric features, the $(r-i)_0$ and the $(u-g)_0$ colors (right panels). The two upper panels of Figure~\ref{feature_imp} show clearly that the SDSS/SEGUE sample do not contain a large number of metal-poor dwarfs in the temperature range that overlap with the majority of giants ($4,900\leq $T$_{eff} \leq 5,500$ K). The selection criteria for SDSS/SEGUE are generally complex \citep{yanny_2009a}. However, the absence of metal-poor dwarfs is exacerbated by the relatively shallow depth of the SDSS/SEGUE dataset, that does not contain stars fainter than $G \simeq 18$ mag. This means dwarfs are generally quite close and so are preferentially selected from the disk, which is much more metal rich on average than the halo.

As a consequence of this, the fraction of true dwarfs misidentified as giants increases drastically with the metallicity in this temperature region, as shown in Figure~\ref{frac_miss}. The majority of true dwarfs with  [Fe/H]$ < -1.5$ are classified as giants by the algorithm. This should be compared to the overall sample, in which the fraction of misidentified dwarfs never exceeds 0.2 (and which is almost constant for metallicity lower than [Fe/H]$=-1.0$). Intriguingly, it is fascinating to note that even in the temperature range including the giants, more than 50\% of true dwarfs (and true giants) are correctly identified between $-1.45 \leq$[Fe/H]$\leq -1.1$. This demonstrates that additional features, not just those relating to temperature and metallicity, are been used by the algorithm to classify dwarfs and giants. It seems reasonable to suppose that these feature relate directly to the surface gravity of the stars, such as the Paschen lines presents in the $i$, $z$ and $G$-bands or the Ca H\&K absorption lines in the $u$, $g$ and $G$-bands \citep[see][]{starkenburg_2017}. 

Our finding partially contradicts \citet{lenz_1998}, who show that it is not possible to simultaneously separate {\it cleanly}
stars by temperature, metallicity, and surface gravity using the SDSS filter set (which are broadly similar to the CFIS and PS1 filters), with the notable exception of A-type stars. However, their analysis did not take into account possible non-linear relations between photometric colors and the relevant stellar parameters. By construction, our RFC accounts for non-linearity in these relations,  allowing us to use photometry to better predict the probability of a star being a dwarf or a giant. We also note that we experimented with different methods to classify dwarfs and giants, including ANNs and a principal component analysis (PCA). We found that the ANN gives similar results to the RFC, but the results were less easy to interpret; the PCA did not produce good results due to its requirement of linearity. 

With the advent of new spectroscopic survey in the northern hemisphere, such as WEAVE and SDSS-V, the number of metal-poor dwarfs with a effective temperature between $4,900\leq $T$_{eff} \leq 5,500$K that have spectra will likely increase. These future data will be excellent training sets to improve the dwarf/giant classification. Indeed, we will discuss later other issues with the training/test sets for which future spectroscopic datasets will likely provide essential improvements.

It is important to keep in mind that the giant sample produced by the algorithm may contain a non negligible fraction of actual metal-poor dwarfs, since dwarfs generally outnumber giants in any survey. However, in the critical temperature range ($4,900\leq $T$_{eff} \leq 5,500$ K), the difference in absolute magnitude between a true dwarf and true giant at the same color is of at least $\simeq 3$ mag, equivalent to an incorrect distance of at least $\simeq 150\%$ (this error being even larger for redder stars). Thus, many of the misidentified dwarfs in our survey can be easily identified by consideration of their Gaia proper motions.

\subsection{Metallicities and distances}

The next step of the algorithm is to determine, independently for each of the two classes, the photometric metallicity and the absolute magnitude (and thus the distance) of each star.

There have been many studies that estimate the photometric parallax of stars, especially MSs \citep{laird_1988,juric_2008,ivezic_2008,ibata_2017b,anders_2019}. This is possible because the MS locus has a well defined color-luminosity relation. Using this property, \citet{juric_2008} derived the distances of 48 million stars present in the SDSS DR8 footprint out to distances of $\sim 20$ kpc using only $r$ and $i$-band photometry. However, this study did not take into account the effects of metallicity, which shifts the luminosity for a given color, more metal rich-stars being brighter than the metal-poor ones (\citealt{laird_1988}, but also see Figure 3 from \citealt{gaiacollaboration_2018b}). Age has less impact on the photometric parallax because its effect is to depopulate the bluer stars while maintaining the shape of the MS locus for the redder stars. 

It is well known that it is possible to derive the photometric metallicity of a star by measuring its UV-excess \citep{wallerstein_1960,wallerstein_1962,sandage_1969}, since metal-poor star have a stronger UV-excess than metal-rich stars. \citet{carney_1979} shows that is is possible to measure the metallicity of a star with a precision of 0.2 dex for stars with photometry better than 0.01 dex in the Johnson $UBV$ filters. More recently \citet{ivezic_2008} showed that it is possible to obtain a similar precision with the $ugr$ SDSS filters, and used it to derive the photometric parallax of 2 million F/G dwarfs up to 8 kpc, where the distance threshold is limited by the precision of the SDSS $u$-band \citep[see also][]{ibata_2017b}.

In this section, we build on this body of literature and present a data-driven method to estimate the metallicity and distances of dwarf stars (Section \ref{sec_dwarf}) and giants (Section \ref{sec_giant}). 

\subsubsection{Dwarf stars} \label{sec_dwarf}

 \begin{figure*}
\centering
  \includegraphics[angle=0, clip, viewport= 5 25 500 810
  , width=14.0cm]{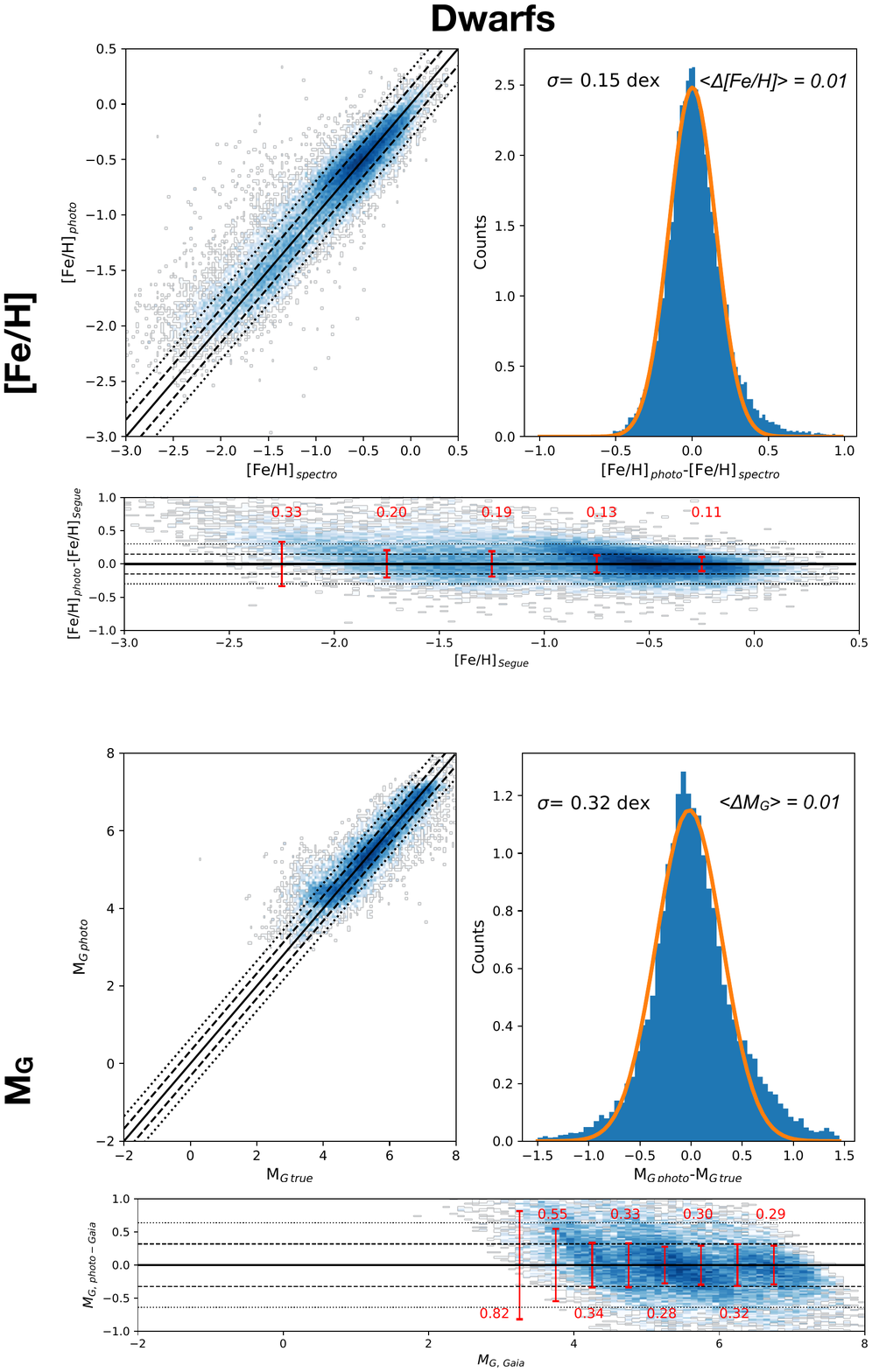}
   \caption{Comparison of the ``true'' and derived metallicity for dwarfs (top panels), and the true and derived absolute magnitudes for dwarfs (bottom panels). The left panels show the true and derived quantities plotted against each other. The one-to-one relation is shown (solid lines), and the dashed and dotted lines correspond respectively to the 1-$\sigma$ and 2-$\sigma$ deviation. The right panels show the distribution of the differences between the true and derived quantities, with a Gaussian fit overlayed. The horizontal panels shows the residue between the quantity predicted by the algorithm and the ``true'' values. The lines are the same than on the left panels. The red error bars show the scatter of the derived parameters at different location.}
\label{residu_dwarf}
\end{figure*}

We first determine the photometric metallicity of the dwarfs before computing the distance of these stars. To determine the distance, we prefer to use the absolute luminosity derived using the Gaia parallaxes, rather than the parallaxes themselves, considering that a large number of Gaia parallaxes are negative due to the stochasticity of the survey \citep{luri_2018a}. By construction, using the absolute magnitude leads to derived distances that are always positive. 

A degeneracy exists between metallicity, luminosity, and colors, especially for stars of high metallicity \citep{lenz_1998}. As pointed out by \citet{ibata_2017b}, neglecting the impact of metallicity on the derived absolute magnitude can lead to important errors, rendering the derived distances invalid. In principle, it should be possible to obtain the metallicity and absolute magnitude of the dwarfs simultaneously, in a single step. However, as described below, only about half of the dwarfs that have good metallicity measurements also have good enough parallax measurements to estimate their absolute magnitudes. In order to use the maximum amount of information available, we decide to use two steps, the first to determine the photometric metallicity and the second to estimate the absolute magnitude.

To evaluate the photometric metallicity of the dwarfs, we construct a set of five independent ANNs, whose inputs are the same colors used for the dwarf - giant classification. Using a set of independent ANNs rather than only one is preferable because it allows us to estimate the systematic errors on the predicted values generated by the algorithm. Using a sub-sample of the training set, we found that 5 independants ANNs gives similar systematic errors that with 10 or 15. Moreover, it also prevents any eventual over-fitting. The five ANNs, constructed using the \textsf{Keras} package \citep{chollet_2015}, have different individual architectures and are composed of between two and five hidden layers. As for the RFC, the training  set  is  used  to  find  the  best  architecture of each ANN with a {\it k-fold} cross validation method with five sub-samples, where we impose that each ANN has an independent architecture. The parameters used to train/test\footnote{As for the dwarf - giant classification, the spectroscopic dwarf dataset is split between a training and test set. However, the results shown in Figures \ref{residu_dwarf} and \ref{residu_giant} are made with the combined dataset to improve their clarity.} the ANNs are the adopted spectroscopic metallicities from the SSPP (\textsf{FeHadop}) and their uncertainties for $\sim 35,000$ dwarfs from the SDSS/SEGUE dataset. The techniques mentioned earlier for estimating metallicity from photometric passbands have typical uncertainties of $\delta$[Fe/H]$\simeq 0.2$ (relative to the spectroscopic measurement). A quality cut is applied on the spectroscopic dataset to only use dwarfs with adopted uncertainities on the spectroscopic metallicity of $\delta$[Fe/H]$_{spectro}\leq 0.2$. We note that this criterion has only a very small impact on the SDSS/SEGUE dwarf catalog, since it removes $\sim 100$ stars.

The loss (or cost) function used to train the ANNs is a modified root mean square function which includes the uncertainties on the metallicity:

\begin{equation}
   \mathcal{L} = \sqrt{\frac{1}{n} \sum_{i=1}^{n}\frac{(y_{true,i}-y_{pred,i})^2}{\delta y_i^2} },
\end{equation}

\noindent where $y_{true,i}$ and $\delta y_i$ are the spectroscopic metallicity and its uncertainty for the $i^{th}$ star, and $y_{pred,i}$ is the corresponding metallicity predicted by the algorithm.

Once each ANN is trained, we define the metallicity of the dwarfs ([Fe/H]$_{Dwarf}$) as the median of the outputs of the five ANNs, and the systematic error as their standard deviation. The difference between the photometric and the spectroscopic metallicity is shown in the two upper panels of Figure \ref{residu_dwarf}, and is $\sigma_{[Fe/H]}= 0.15$ dex. This is a moderate improvement on the method of \citet{ibata_2017b} ($\sigma_{[Fe/H]}= 0.20$ dex). Note that the residuals do not show any significant trend with the metallicity, except for the most metal-poor stars ([Fe/H] $< -2$) for which the photometric metallicity tend to be higher than the spectroscopic measurement. We remark that the residuals increase with decreasing metallicity, indicating that the predicted photometric metallicities are less reliable for stars with metallicity lower than [Fe/H]$<-2.0$. This will be partly due to the lower number of stars present in the training set at this metallicity than at higher metallicity. The average uncertainty on the spectroscopic metallicity of the dwarfs is $\delta$[Fe/H]$=0.04$ and the systematic error is $\delta $[Fe/H]$_{Dwarf,sys}=0.02$. Thus, most of the scatter on our metallicity measurement is due to the intrinsic apparent color variation of the dwarfs of similar metallicity. 

Once the metallicity is determined, it is used in combination with the colors to estimate the absolute magnitude of each dwarf, and therefore its distance. As for the metallicity, a set of five ANNs is constructed, where the inputs are the same colors used previously in addition to the derived metallicity. The output is the absolute magnitude in a given band. We decide to use the Gaia $G$ band as a reference since this is the filter with the lowest photometric uncertainty at a given magnitude. 

The absolute magnitude of the dwarfs in the SDSS/SEGUE catalog are computed from Gaia parallaxes ($\varpi$) according to
\begin{equation}
M_{G_{gaia}}= G_0 + 5 + 5\log_{10}(\varpi/1000).
\label{eq_MG}
\end{equation}

It is worth noting that Gaia tends to underestimate the parallaxes and we therefore correct all parallaxes by a global offset of $\varpi_0=0.029$ mas, as suggested by \citet{lindegren_2018}. \citet{luri_2018a} show that the inversion of the parallax to obtain the distance (and so the absolute magnitude), is only valid for the stars with low relative parallax uncertainties, typically $\varpi/\delta\varpi \geq 5$ (a relative precision of $\leq 20\%$). In these cases, the probability distribution function (PDF) of the absolute magnitude can be approximated by a Gaussian centered on $M_{G_{gaia}}$ with a width $\delta M_{G_{gaia}}$, such that

\begin{equation}
   \delta M_{G_{gaia}}=\delta G + \left|\frac{\delta \varpi}{\varpi \ln(10)}\right| . 
 \label{eq_dMG}
\end{equation}

A quality cut is performed on the SDSS/SEGUE dwarfs to keep only those stars with a relative Gaia parallax measurement better than 20\%. With this criterion, the mean relative parallax precision of the spectroscopic dwarf sample is $\sim 10\%$, corresponding to an average uncertainty on the absolute magnitude of $\delta M_{G_{gaia}}=0.22$ mag. The spectroscopic dwarf dataset used to train/test the set of ANNs is composed of $18,930$ stars, and is a good representation of the distribution of metallicities and absolute magnitudes in the initial dataset (covering a range of metallicity of $-3.0<$[Fe/H]$_{spectro}<0.5$ dex, a range in absolute magnitude $3 < M_G < 7.5$, and $\simeq 600$ stars with [Fe/H]$_{spectro} < -2.0$ with good parallax precision). 

The set of five ANNs used to derive the absolute magnitude have a different structure than the set used for the estimation of the metallicity, with four or five hidden layers and a higher number of neurons per layer than previously. However, the loss function is the same, where $y_{true}$, $\delta y$ and $y_{pred}$ now correspond to $M_{G_{gaia}}$, $\delta M_{G_{gaia}}$ and $M_{G_{pred}}$, respectively. We define the predicted absolute magnitude in the $G$-band of the dwarfs (M$_{G_{Dwarf}}$) as the median of the outputs of the five ANNs, and the systematic error ($\delta$ M$_{G_{Dwarf},sys}$) as their standard deviation. As illustrated in the lower panels of Figure \ref{residu_dwarf}, the predicted absolute magnitude shows a scatter of $\sigma_{M_G}=0.32$ mag compared to the absolute magnitude computed from the Gaia parallaxes. This corresponds to a relative precision on distance of 15\%, very similar to the precision found by \citet{ivezic_2008}. It is worth noting that the scatter is almost constant over the range of absolute magnitude, except around $M_{G_{gaia}} \simeq 4.2$, where a few stars tend to have a higher predicted absolute magnitude than observed. These stars are probably young stars ($< 5$ Gyr) on the main-sequence turn-off (MSTO). Since CFIS observes at high galactic latitude ($|b|> 18 \degr$ ), their number is negligible, and we expect that this under-estimation of the luminosity of younger stars should have a negligible impact on statistical studies of the distance distribution.

\subsubsection{Giant stars} \label{sec_giant}

 \begin{figure*}
\centering
  \includegraphics[angle=0, clip, viewport= 5 0 550 845
  , width=14.0cm]{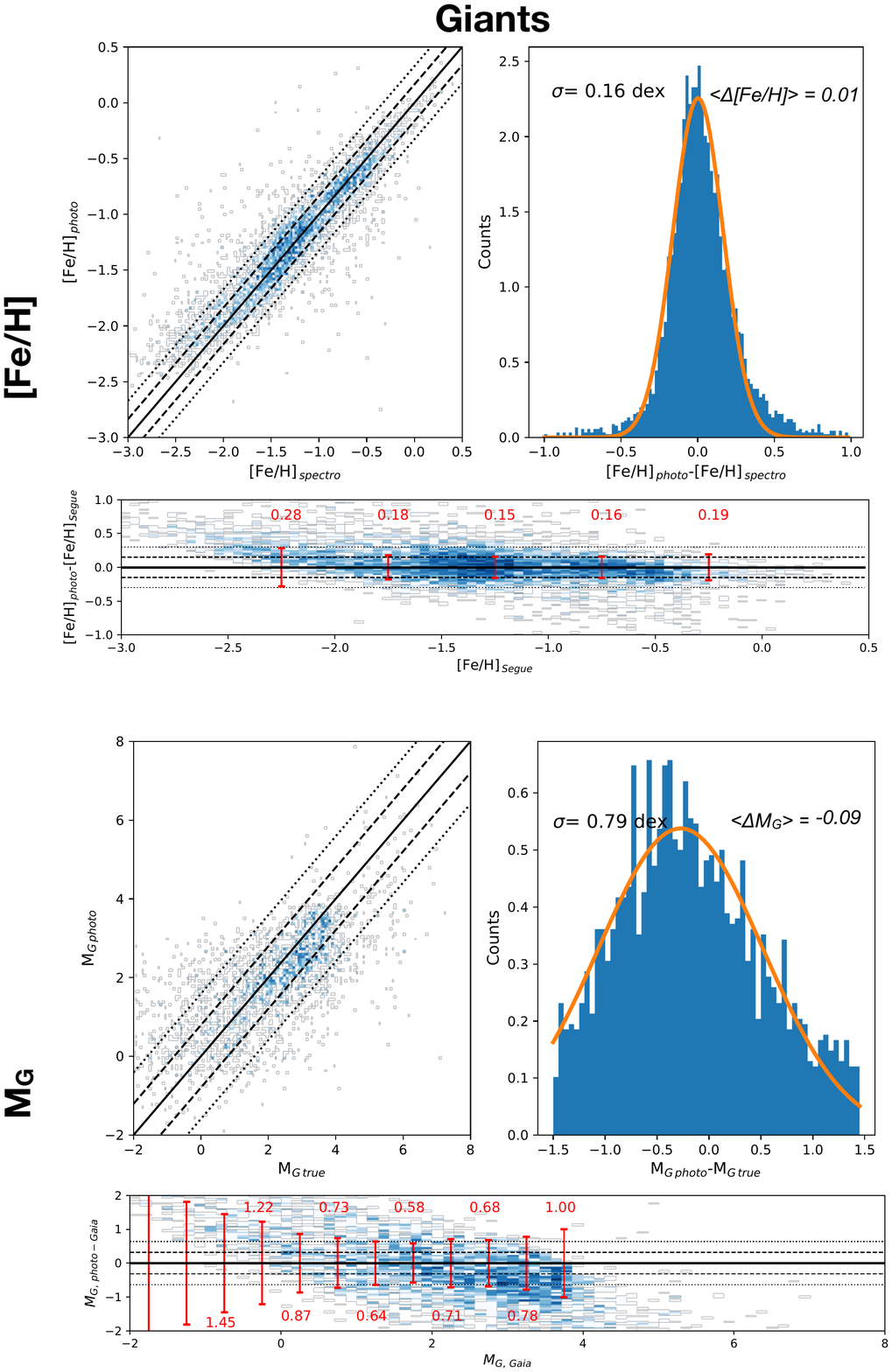}
   \caption{Same as Figure \ref{residu_dwarf} but for giants.}
\label{residu_giant}
\end{figure*}

The method to derive the metallicity and the absolute magnitude is similar for the giants as for the dwarfs, with a first set of ANNs to derive the metallicity, and second set of ANNs to estimate the absolute magnitude. The architecture of the two sets of ANNs are exactly the same as for the dwarfs.

The adopted metallicities and the uncertainties for the spectroscopic giants are used to train/test the first set of ANNs. Again, we apply a quality cut on the metallicity that uses only the $5,670$ giants with a metallicity precision better than $\delta$[Fe/H]$_{spectro}\leq 0.2$ (this quality cut removes less than $0.5\%$ of the initial giant sample).
The procedure used is exactly the same as for the dwarfs, and the predicted metallicity of the giants ([Fe/H]$_{Giant}$) is equal to the median of the five ANNs outputs, and the systematic errors ($\delta$[Fe/H]$_{Giant,sys}$) is the standard deviation of these outputs. As shown on the two top panels of Figure \ref{residu_giant}, the residual between the photometric and spectroscopic metallicity is $\sigma_{[Fe/H]}= 0.16$ dex and does not show any trend with metallicity, except for stars with [Fe/H]<-2.0 as for the dwarfs. This is a very significant improvement over previous studies, where the giants were not treated separately from the dwarfs, leading to an overestimation of their metallicity by [Fe/H]$= 0.16$ dex \citep{ibata_2017b}.

In contrast to the dwarfs, it is not possible to keep only the giants with a relative Gaia parallax accuracy $\lesssim 20$\%. At a similar luminosity, the giants are more distant than the dwarfs, which leads to a higher uncertainty in their parallax. Adopting the same selection as for the dwarfs creates a dataset of only $\simeq 1,000$ stars, whose overall metallicity distribution is very different from the overall metallicity distribution of the giants. Indeed, the large majority of these stars ($\sim 670$) have a metallicity higher than [Fe/H]$>-1$ while 3/4 of the overall spectroscopic giants dataset has a metallicity lower than [Fe/H]$<-1$ with a peak around [Fe/H]$\simeq -1.4$ (see the lower panel of Figure \ref{MDF}). In addition, more than 60\% of the giants with a parallax accuracy of $\lesssim 20$\% seem to be sub-giant stars, misidentified by our dwarf - giant classifier. 

About 2/3 of the spectroscopic giants have a relative precision on their parallax measurement higher than 20\%. Therefore, the PDF of their absolute magnitude cannot be approximated by a Gaussian, as was the case for the dwarfs. As shown by \citet{luri_2018a}, their PDF is asymmetric and the maximum is not centered on the ``true'' absolute magnitude. Using the maximum of the PDF in these cases leads to an underestimate of the distance, and therefore underestimates the absolute magnitude. Depending on the relative parallax precision, the ``true'' absolute magnitude can be more than 1-$\sigma$ away from the maximum likelihood value. 

In principle, we could perform a {\it data augmentation} of the spectroscopic giants dataset to take into account the uncertainties on the different parameters, especially the parallax. This can be done by Monte-Carlo sampling the spectroscopic giants dataset in a range of 2.5 - 3 $\sigma$ around the maximum likelihood {\it of the observables}. In order not to add any bias, this distribution should be symmetric around the maximum likelihood of the different parameters. However, to keep a physical value of absolute magnitude, the parallaxes should be positive. This leads to a dataset composed of stars with $\varpi-2.5*\delta \varpi > 0$, reducing drastically the number of stars used to train the model, which also biases the sample to much more metal-rich stars, similar to the effect of cutting on the parallax uncertainties described above.

For these reasons, we use all the spectroscopic giants that have a positive parallax measurement. Following \citet{hogg_2018}, we apply a quality selection on the parallax to keep only giants with an uncertainty on the parallax $\delta\varpi<0.1$ mas, so that we are not dominated by stars with extremely poor measurements. The final dataset used to train/test the set of ANNs is composed of $3,497$ giants. However, we found that for this giant sample, the Gaia parallaxes have to be corrected by an offset of $\varpi_0=0.033$ mas in order to obtain reasonable distances for the globular clusters (see Section \ref{sec_GC}). This offset is slightly higher than for the dwarfs (of $\varpi_0=0.029$ mas), but lower than the offset of $\varpi_0=0.048$ mas found by \citet{hogg_2018} for red giant stars.

Due to a lack of precise parallaxes, the exact PDF (and so the uncertainties) of the absolute magnitude (calculated using Equation \ref{eq_MG}) cannot be known for most of the giants without adopting a prior on the density distribution of the giants in the Milky Way. We cannot, therefore, estimate the uncertainties on the absolute magnitude using Equation \ref{eq_dMG}. For this reason, the loss function used as a metric for the ANNs is a standard root mean square that does not take into account the uncertainties on the absolute magnitude. 

The scatter on the absolute magnitude shown in the left panel of Figure \ref{residu_giant} is $\sigma_{M_G} = 0.79$ dex, with a bias of $\sim -0.28$ dex, similar to the median of the residuals (Me$(\Delta M_G)=-0.28$). This bias is a consequence of the underestimation of the distance to the giants using the values given by the maximum of the PDF. The mean residual is larger ($<\Delta M_G>=-0.09$), because it is influenced by those stars with the largest residuals, a result of the limited statistical sample.

 Indeed, the Gaia parallaxes of these stars are inaccurate, and distances obtained by inverting the parallax tend to underestimated the true distance of these stars \citep{luri_2018a}. Thus the absolute magnitudes used to train the relation are inaccurate, and are likely underestimate. It is therefore not possible to use the scatter between the predicted absolute magnitude and the absolute magnitude obtain from the Gaia parallax to verify that the predicted distances are correct with this method. However, as we will show in Section \ref{sec_GC} using the globular clusters present in the CFIS footprint, the distances of the giants thus determined gives good estimates of their real distances, despite the lack of precision of the absolute magnitudes used to train the relation.

\section{Performance of the algorithm} \label{verif}

\subsection{Verification from SDSS/SEGUE}

\begin{figure}
\centering
  \includegraphics[angle=0, clip, width=7.5cm]{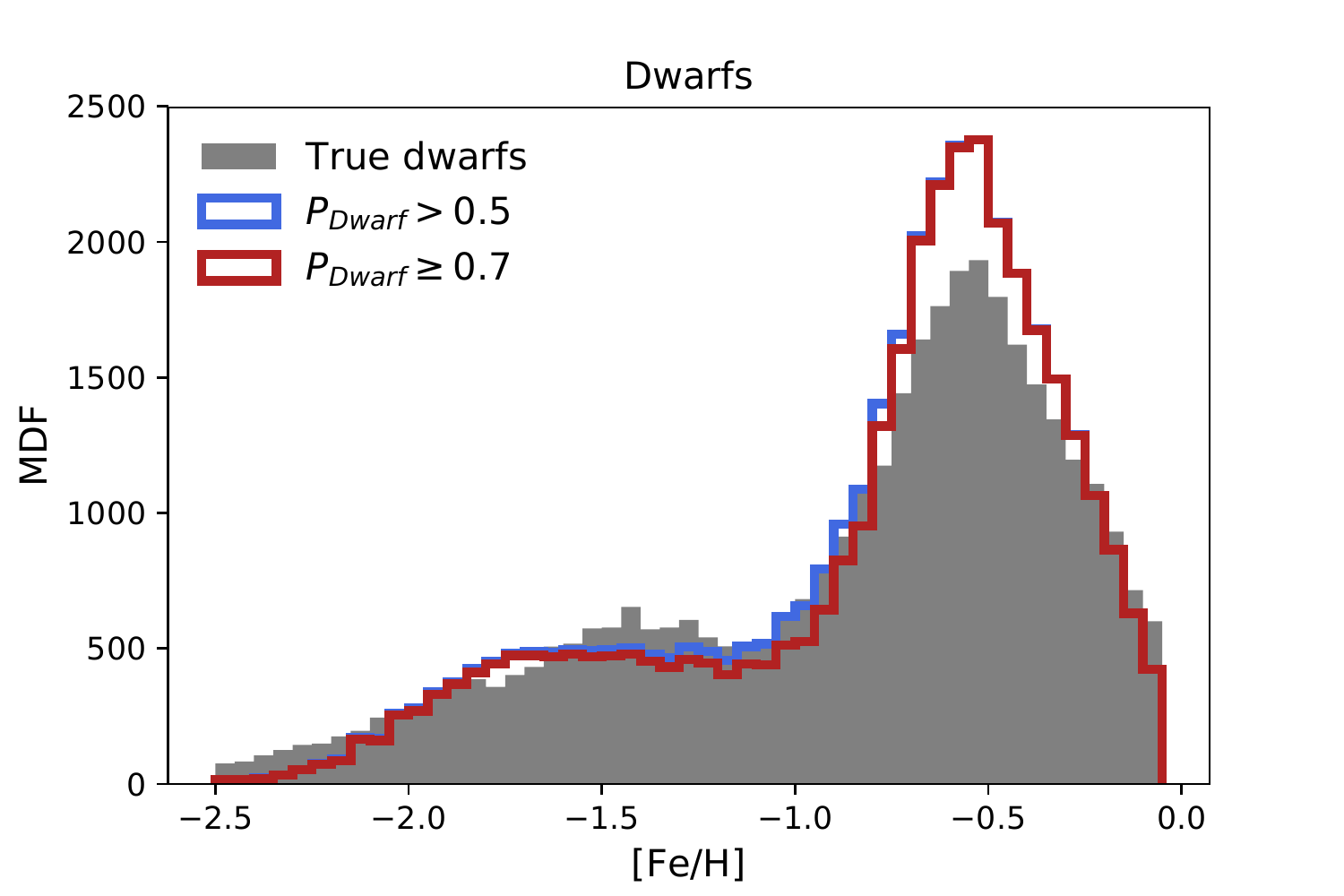}
    \includegraphics[angle=0, clip, width=7.5cm]{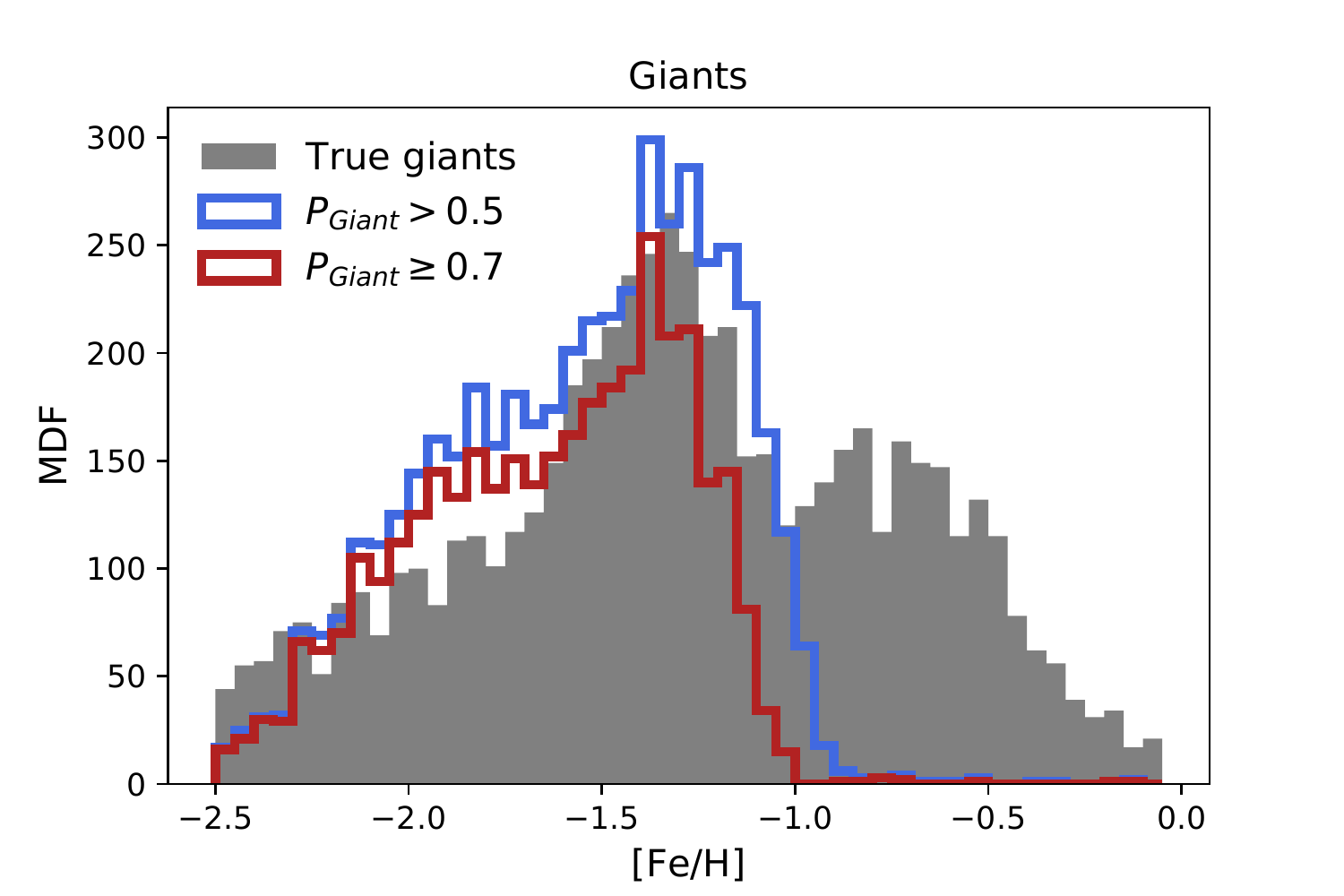}
   \caption{Metallicity distribution function (MDF) of the dwarfs (upper panel) and giants (lower panel). The gray histograms are the spectroscopic MDFs from the SDSS/SEGUE catalog using the adopted metallicity from the SSPP. The blue and red histograms are the predicted photometric MDFs for the stars, where blue and red lines correspond to stars with a confidence of being a dwarf/giant of 0.5 and 0.7, respectively. The excess of dwarfs around [Fe/H]$\simeq -0.5$ and the depletion of giants more metal-rich than [Fe/H]$=-1.0$ are the consequence of the mis-classification of actual metal-rich giants by the algorithm.}
\label{MDF}
\end{figure}

\begin{figure*}
\centering
  \includegraphics[angle=0, clip,viewport= 0 0 1015 550, width=17cm]{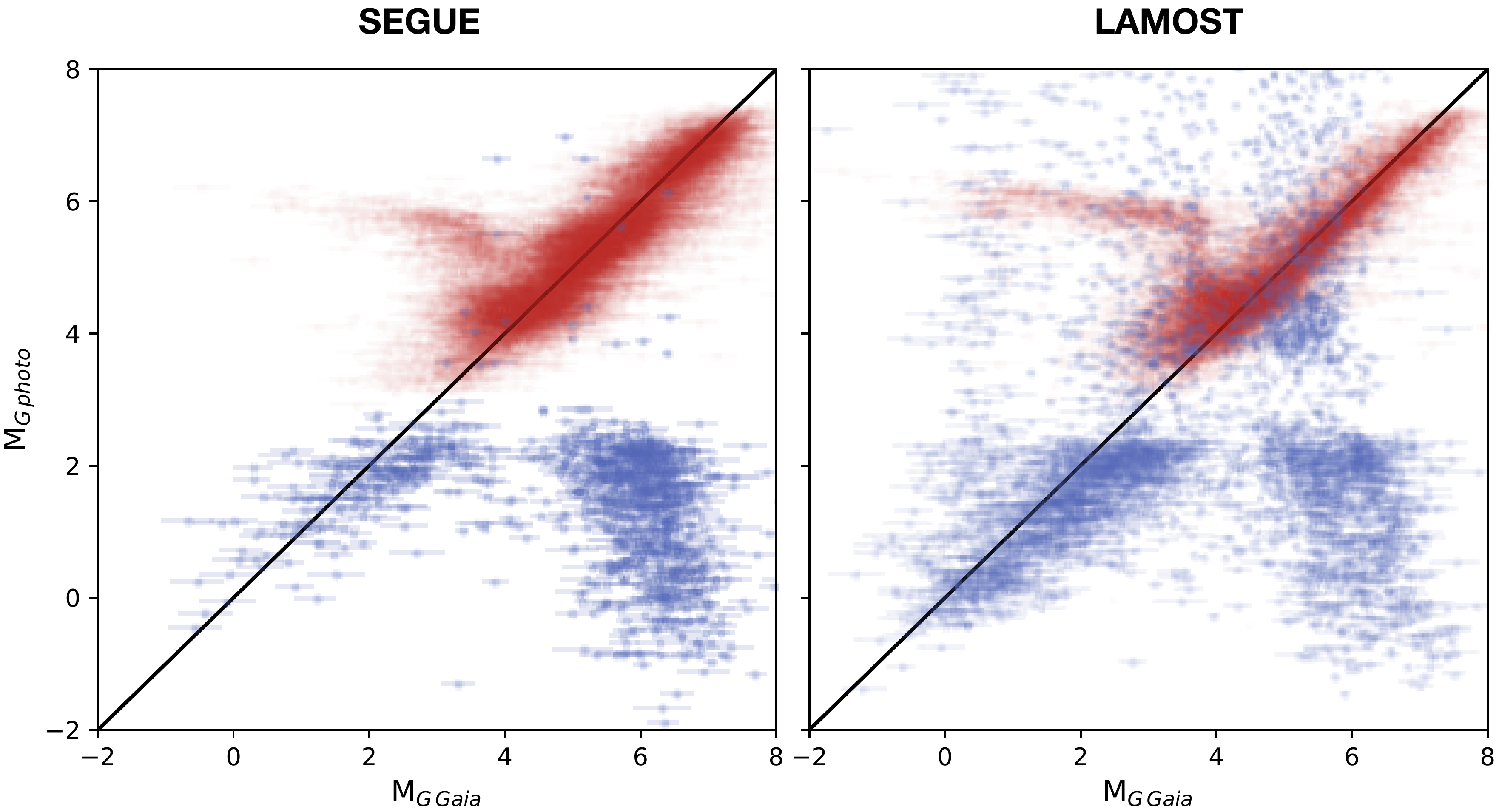}
   \caption{Distribution of the absolute magnitude computed from the Gaia parallax (x-axis) against the absolute magnitude predicted by the algorithm (y-axis), for stars with $\varpi / \delta \varpi \ge 5$. The left panel is for stars from SDSS/SEGUE dataset, and the right panel is for stars from the LAMOST dataset. The stars predicted as dwarfs are in red and the stars predicted as giants are in blue.}
\label{MG_distri}
\end{figure*}

We now apply the algorithm to the full SDSS/SEGUE dataset for stars with a spectroscopic $SNR \geq 25$. The photometric metallicity distribution function (MDF) from the stars classified by the algorithm as dwarfs ($P_{Dwarf} \geq 0.5$) and giants ($P_{Giant} \geq 0.5$) are compared to the spectroscopic metallicity distribution of the dwarfs and giants for the SDSS/SEGUE in Figure \ref{MDF}. 

The expected and predicted MDFs for dwarfs and giants in Figure~\ref{MDF} are generally similar, especially for the dwarfs. For the giants, the agreement at the metal-poor end is good, but the number of giants predicted to have [Fe/H]$> -1.0$ falls to zero. This is a direct consequence of the mis-classification of the most metal-rich giants, as discussed in Section \ref{sec_DG}. Further, these mis-classified giants are the origin of the excess of dwarfs around [Fe/H]$\simeq -0.5$ compared to the number expected from the spectroscopy. 

If we consider only stars classified as dwarfs/giants with high confidence ($P_{Dwarf/Giant} \geq 0.7$; red lines in Figure~\ref{MDF}),  the MDF of the dwarfs is almost unchanged. This means that the large majority of dwarfs are classified by the algorithm with high confidence. For the giants, the MDF of stars classified with high confidence decreases sharply  at [Fe/H]=-1.2, instead of [Fe/H]=-1.0 for the stars with $P_{Giant} \geq 0.5$. {\it Thus, the classification of giants is more uncertain at high metallicities than at lower metallicities.} Again, this is another manifestation of the difficulty of the algorithm in identifying more metal-rich giants. The over-predicted number of giants around [Fe/H]=-1.8 is a consequence of the trend in the photometric metallicity relation which tends to over-estimate the metallicity of stars with [Fe/H]$<-2.0$, as explained in section \ref{sec_giant}.

The left panel of Figure \ref{MG_distri} compares the absolute magnitude expected using the Gaia parallaxes ($M_{G,Gaia}$) with the absolute magnitude predicted by the algorithm ($M_{G,photo}$) for stars in SDSS/SEGUE with $\varpi/\delta \varpi \geq 5$. The stars predicted as dwarfs and giants by the algorithm are shown in red and blue, respectively. As expected, the predicted absolute magnitude of the large majority (more than 97\%)  of the stars identified as dwarfs is similar to the Gaia measurements. However, a small population of dwarfs have predicted absolute magnitudes significantly lower than observed (around $M_{G,Gaia} \sim 3$). This population corresponds to metal-rich giants/sub-giants that have been misidentified as dwarfs. For those stars identified as giants by the algorithm that are actually giants, the agreement between the predicted and actual magnitudes is good. However, all those stars identified as giants that are actually dwarfs are estimated to be too bright. 

 It is not surprising, given the criterion imposed on the relative precision of the parallax, that the fraction of contaminant dwarfs is much higher than the $\simeq 30\%$ measured previously. Indeed, the Gaia catalog is limited at a magnitude of $G = 21$ and since dwarfs are intrinsically less bright than giants, dwarfs in this catalog are on average closer than giants. Since the parallax method is more precise for the stars at shorter distances, it is natural that most of the stars with relative parallax measurements better than 20\% are actual dwarfs. Nevertheless, it is very interesting to see that the actual giants have predicted magnitudes similar to Gaia's. Moreover, the scatter is only slightly larger than for the dwarfs, despite the fact that the data used to train the set of ANNs are less accurate for giants than for dwarfs.

\subsection{Verification with LAMOST}

 \begin{figure}
\centering
  \includegraphics[angle=0, clip,viewport= 0 0 780 770, width=7.5cm]{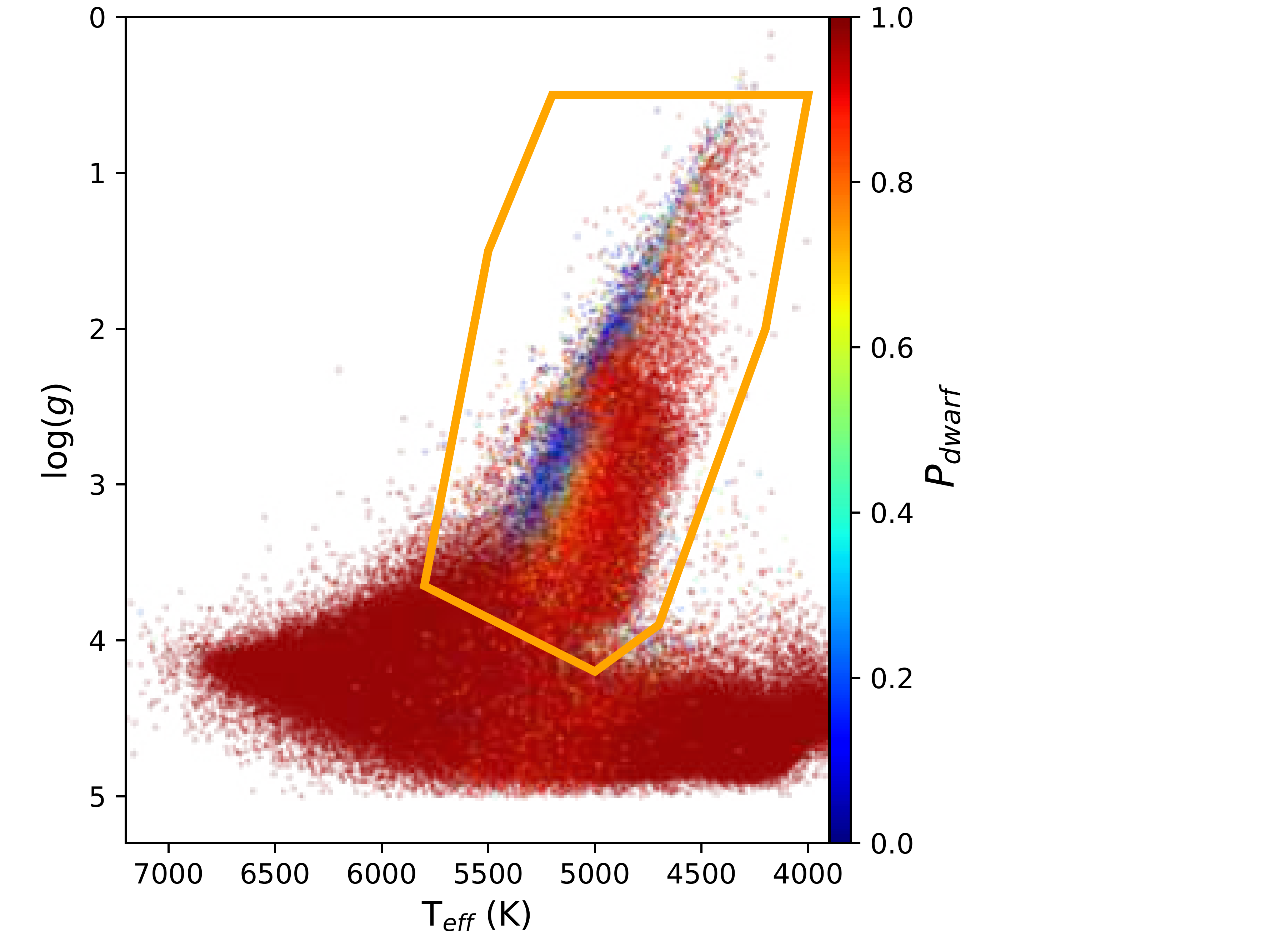}
   \caption{Same as Figure \ref{HR} but for the LAMOST dataset.}
\label{HR_LAMOST}
\end{figure}

Approximately $480,000$ stars from the LAMOST DR3 catalog are present in the CFIS footprint\footnote{Contrary than for SDSS/SEGUE, we do not limit our analysis to  stars with a higher signal-to-noise ratio.}. While this is 10 times more than for SDSS/SEGUE, most LAMOST stars are metal-rich. Less than 3,000 LAMOST dwarfs have a metallicity of [Fe/H] $\leq -1.5$, and their metallicity accuracy is lower than for the corresponding metallicity range in the SDSS/SEGUE dataset. Scientifically, we are more interested in distant stars, at the faint end of the Gaia catalogue, so we focused the training of the algorithm on the SDSS/SEGUE dataset instead of the LAMOST dataset. However, this means that the LAMOST dataset can be used to independently test and validate our algorithm. 

\begin{figure}
\centering
  \includegraphics[angle=0, clip,viewport= 0 0 1025 690, width=8.5cm]{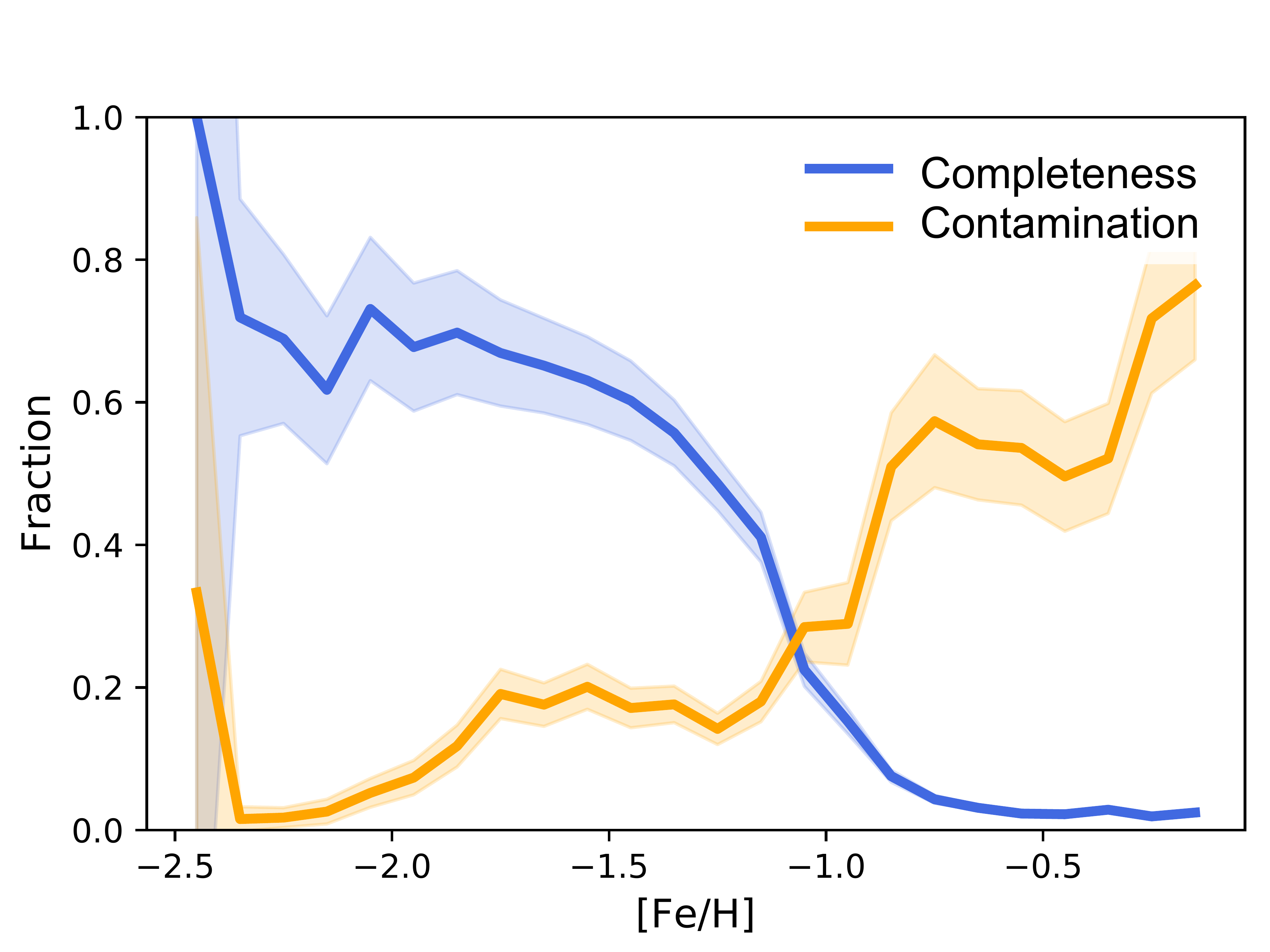}
   \caption{Completeness (blue) and contamination (orange) fraction of the stars from the LAMOST dataset classified as giants by the algorithm as function of the spectroscopic metallicity.}
\label{conta_LAMOST}
\end{figure}

Figure \ref{HR_LAMOST} shows the {\it Kiel} diagram of LAMOST stars color-coded by the probability of being a dwarf according to  our algorithm. Hotter giants are generally classified correctly by the algorithm, but a large number of giants are mis-classified. The completeness and contamination fraction of LAMOST giants is shown in Figure \ref{conta_LAMOST}. Nearly all the mis-classified giants visible in Figure~\ref{HR_LAMOST} have [Fe/H] $>-1.0$, and contamination starts to increase at [Fe/H] $>-1.3$, demonstrating consistency with the SDSS/SEGUE sample that was used to train the algorithm.

\begin{figure}
\centering
  \includegraphics[angle=0, clip,viewport= 0 0 485 770, width=8.5cm]{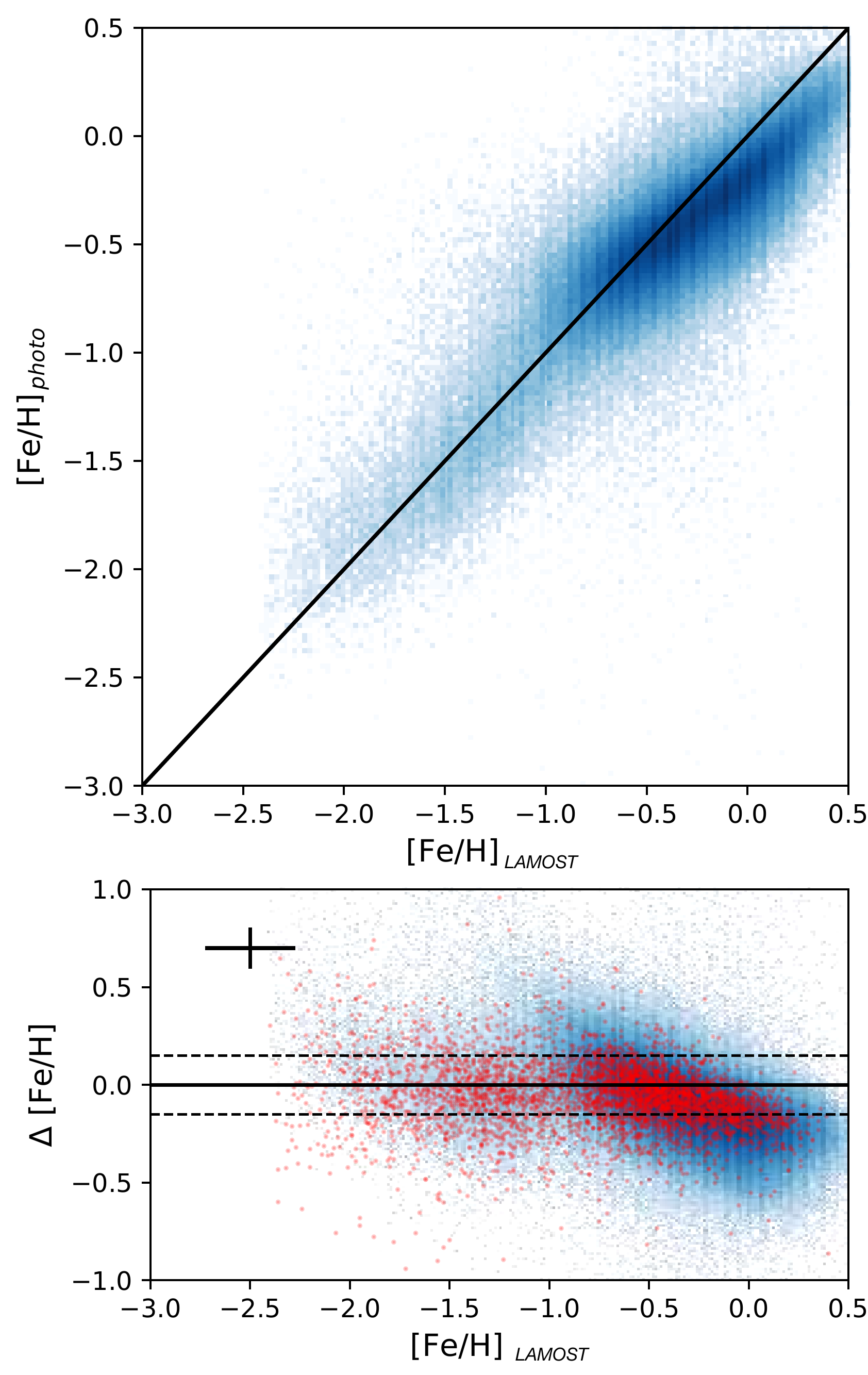}
   \caption{{\it Higher panel:} Comparison of the ``true'' and derived metallicity for all stars in the LAMOST dataset. {\it Lower panel:} residual of the photometric metallicity against the LAMOST value in blue and in red the residual of the SDSS/SEGUE metallicity value from the SSPP against the LAMOST values. The uncertainties on the metallicity values given by LAMOST and the average residual of the photometric metallcity are respectively shown by the horizontal and vertical errorbar. The dotted line shows the 1-$\sigma$ of the giants ($\sigma =0.16$) relation of the photometric metallicity determined previously.}
\label{feh_LAMOST}
\end{figure}

The right panel of Figure \ref{MG_distri} shows that the distribution of the predicted absolute magnitude for the LAMOST stars with $\varpi/\delta \varpi \geq 5$ is similar to the corresponding distribution for the SDSS/SEGUE stars (left panel). We note that the relative fraction of actual dwarfs mis-identified as giants (bottom right corner) is smaller than for the SDSS/SEGUE dataset. This is because the LAMOST dataset is intrinsically brighter than the SDSS/SEGUE dataset. Indeed, since LAMOST stars are $\sim 2$ magnitudes brighter than SDSS/SEGUE, the giants are on average closer. Thus the number of giants that have a relative parallaxes precision better than 20\% is higher in the LAMOST dataset. 

The number of stars present in the horizontal feature between $0 \leq M_{G,\, Gaia} \leq 4$ is higher in the LAMOST dataset than for SDSS/SEGUE. These stars correspond to actual giants mis-classified as dwarfs. The higher number of these mis-classified stars is higher in the LAMOST dataset than in SDSS/SEGUE because the number of metal-rich giants ([FeH]>-1.0) is higher in this first one and, as mentioned earlier, the dwarfs/giants classification does not work well for these stars.

Figure \ref{feh_LAMOST} shows that the predicted metallicities for LAMOST stars are generally consistent with the spectroscopic metallicities obtained by the LAMOST pipeline. However, the predicted metallicity seems to be slightly under-estimated for stars with [Fe/H]$\geq -0.5$. Interestingly, we trace this to a systematic difference between the spectroscopic metallicity  for the $\sim 4,500$ stars in common between the SDSS/SEGUE and LAMOST datasets, as show by the red dots on the lower panel of Figure \ref{feh_LAMOST}. Since the metallicity is calibrated on the \textsf{FeHadop} metallicity from SDSS/SEGUE, it is not surprising to see a similar trend in the predicted photometric metallicity. 

Based on this comparison, we conclude that our method, applied to the LAMOST dataset, demonstrates the same behaviours, agreements and biases as found using the SDSS/SEGUE dataset.

\subsection{Distant Galactic satellites} \label{sec_GC}

\begin{figure*}
\centering
  \includegraphics[angle=0, clip,  viewport= 0 0 460 845, width=11.3cm]{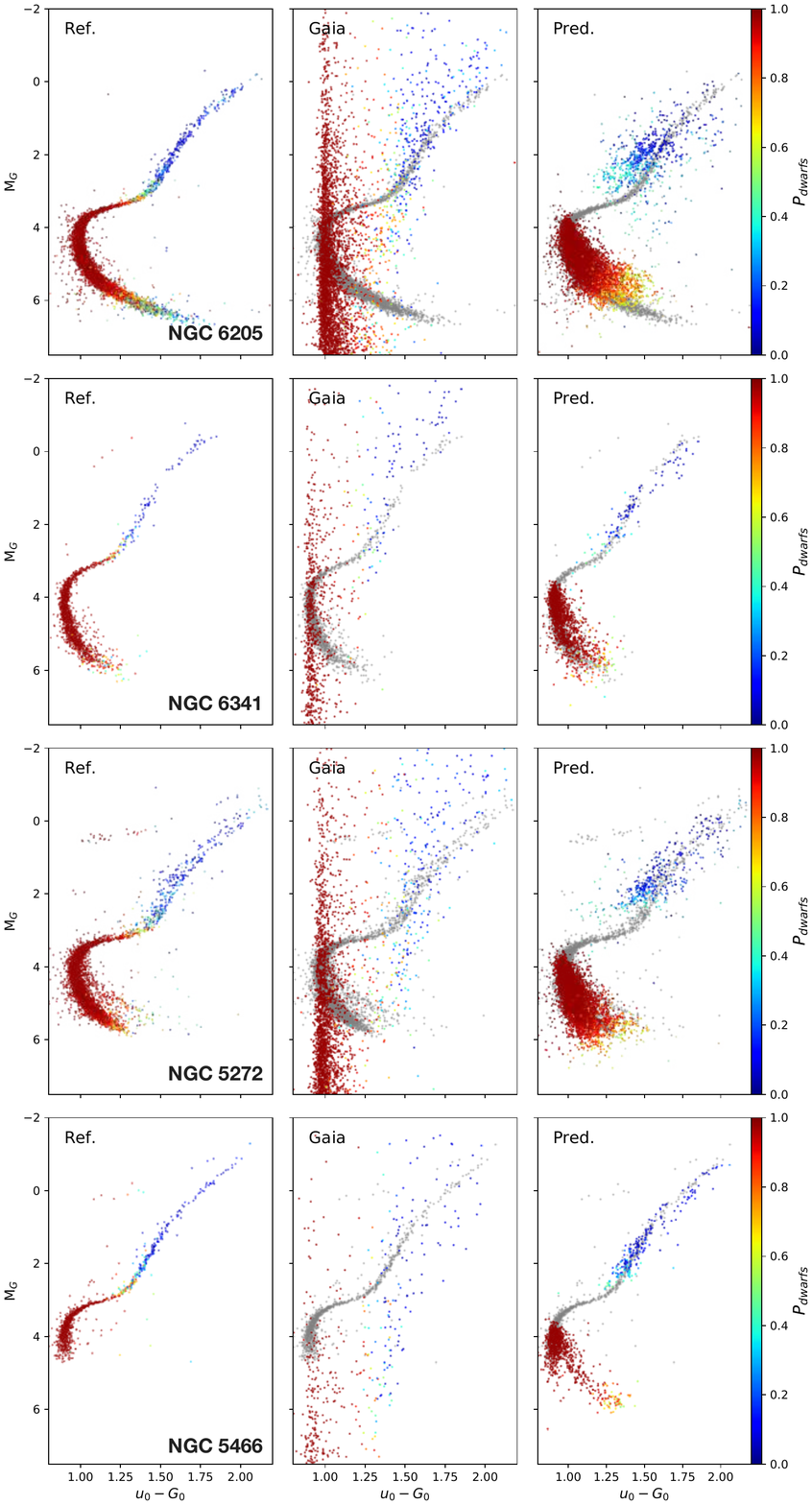}
   \caption{Color-magnitude diagrams of four globular clusters (GCs). Stars are color-coded by the probability of them being a dwarf star. The left panel shows the reference CMD of the cluster, where the absolute $G-$band magnitude is calculated using the distance given in Table \ref{table_GCs}. The middle panel shows the CMD whose the absolute $G-$band magnitude is computed using the $Gaia$ parallax measurement, following Equation \ref{eq_MG}. The right panel shows the CMD where the absolute $G-$band magnitude is calculated using our algorithm. The gray CMD in the middle and right panels is the reference CMD, and is included for easy comparison.}
\label{CMD_GCs}
\end{figure*}

\begin{table*}[ht]
 \centering
  \caption{Predicted mean distances and metallicities for dwarfs and giants in the 4 globular clusters in Figure \ref{CMD_GCs} as derived by our algorithm, compared to literature values for the clusters. Literature values from: \citet{harris_2010} (1), \citet{deras_2019} (2) and \citet{hernitschek_2019} (3)}
  \label{table_GCs}
  \begin{tabular}{@{}l|ccr||ccc@{}}
  \hline
   Name & $D_{dwarfs}$ (kpc) & $D_{giants}$ (kpc) & $D_{ref}$ (kpc) & [Fe/H]$_{dwarfs}$ & [Fe/H]$_{giants}$ & [Fe/H]$_{ref}$  \\
    \hline
NGC6205 & $8.1 \pm{1.4}$ & $7.8 \pm{1.7}$ & $7.1 \pm{0.1}$ (2) & $-1.70\pm{0.3}$ & $-1.50 \pm{0.16}$ & $-1.53$ (1)\\
NGC6341 & $8.9 \pm{1.4}$ & $8.3 \pm{1.1}$ & $8.3 \pm{0.2}$ (1)& $-2.00\pm{0.37}$ & $-2.18 \pm{0.16}$ & $-2.31$ (1)\\
NGC5272 & $10.2 \pm{1.9}$ & $10.9 \pm{2.7}$ & $10.48 \pm{0.07}$ (3) & $-1.8\pm{0.37}$ & $-1.56 \pm{0.18}$ & $-1.50$ (1)\\
NGC5466 & $14.7 \pm{2.4}$ & $16.0 \pm{2.5}$ & $15.76 \pm{0.14}$ (3) & $-1.83\pm{0.40}$ & $-1.88 \pm{0.14}$ & $-1.98$ (1)\\
\hline
\end{tabular}
\end{table*}

\begin{figure*}
\centering
  \includegraphics[angle=0, clip, width=15cm]{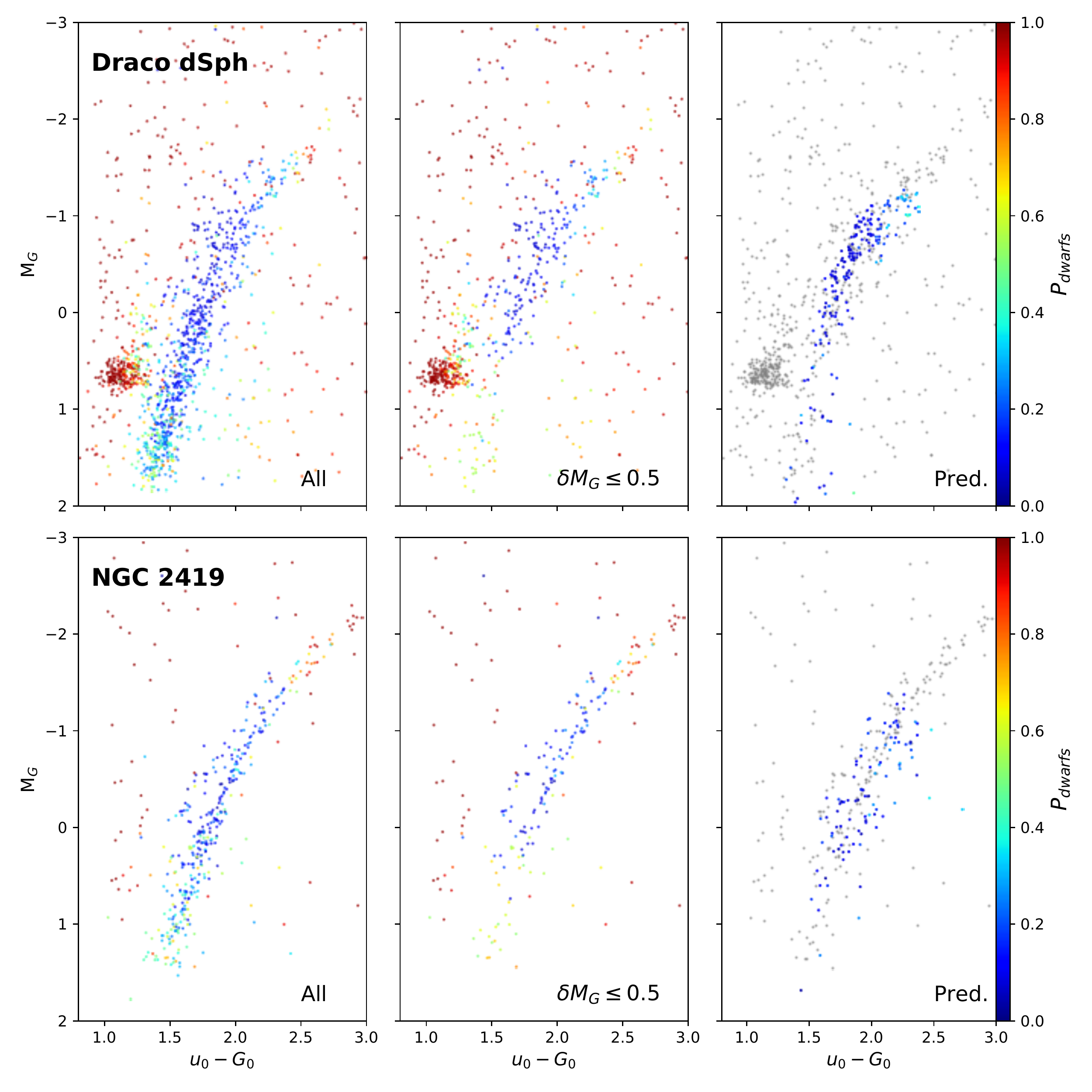}
   \caption{Color-magnitude diagrams of the Draco dSph (top) and of NGC 2419 (bottom). The left panel shows the reference CMD of the object (using all stars in the field) where the absolute $G-$band magnitude is calculated using the literature distance. The middle panels show the CMD of the stars with good precision on their intrinsic absolute magnitude ($\delta M_G \leq 0.5$). The right panel shows the CMD where the absolute $G-$band magnitude is calculated using our algorithm. }
\label{draco_2419}
\end{figure*}

Several Galactic satellites with known distances are included in the CFIS dataset, and present another opportunity to independently test our algorithm. We first examine the four closest globular clusters present in the CFIS footprint, NGC 6205, NGC 6341, NGC 5272 and NGC 5466, where both dwarfs and the giants are detected. These are located between 7 and 16 kpc from us (see Table \ref{table_GCs}). 

For this analysis, we only consider stars at more than 3, 5, 2.5 and 1 half-light radius for NGC 6205, 6341, 5272 and 5466, respectively. This removes the (significant) effect of crowding on the input photometry. We remove obvious foreground contamination by selecting stars using their Gaia proper motions. The absolute color-magnitude diagrams (CMD) of these GCs are shown in Figure \ref{CMD_GCs}. The left panels show the CMDs using the distances found in the literature (fourth column of Table \ref{table_GCs}). Each star is color-coded by the probability that it is a dwarf according to our algorithm. 

The color-coding in the left panels of Figure~\ref{CMD_GCs} reveals that our algorithm generally identifies the giants correctly. A noticeable exception is for horizontal branch stars, which are not present in significant numbers in our training/test sets. In addition, the large majority of the dwarfs in each GC are also correctly identified. The exception to this is for the faintest stars in NGC 6205, where the fraction of dwarfs misclassified as giants increases at the faintest magnitudes. This is likely a direct consequence of the misidentification of metal-poor dwarfs in the temperature range occupied by giants, and discussed at length in Section \ref{sec_DG}.

The middle panel of Figure \ref{CMD_GCs} show the CMDs of the clusters where the absolute $G$-band magnitude is computed directly from the Gaia parallax using Equation \ref{eq_MG}. It is clear that this method is inadequate for these objects, since the uncertainties on the Gaia parallaxes for stars more distant than a few kiloparsecs are generally prohibitively large. {\it We stress that it is this fact that motivated the development of this algorithm in the first place.} The right panels of Figure \ref{CMD_GCs} show the CMDs where the absolute magnitude of the stars is computed via our algorithm. It is clear that the derived CMDs are much better than for the middle panels, and are reasonably close to the reference CMDs for each cluster. We note that the absence of any stars on the sub-giant branch in the CMDs in the right panel is a direct consequence of the preference of our algorithm to define those stars as dwarfs. 

To verify that our algorithm predicts correct metallicities and distances for dwarfs and giants, we compare the mean distance and metallicity of the four GCs according to our algorithm to the value found in the literature. We select stars that have a probability to be a dwarf or a giant of more than 0.7. For the dwarf samples, we require stars to have an intrinsic absolute magnitude larger than 3.7, to remove the impact of the sub-giants. The mean distances and metallicities for dwarfs and giants for each cluster are listed in Table \ref{table_GCs}. For all the parameters, the derived values are within 1-$\sigma$ of the literature values for the clusters. 

It is important to note that the lower precision on the distance estimated for the four globular clusters by our method is of 26\% using the giants and of 18\% using the dwarfs, validating our method to estimate the absolute magnitude of those stars. One could notice that the mean metallicities obtained for the dwarfs for NGC 6205 and NGC 5272 are $\simeq 0.2-0.25$ dex lower than listed in the literature, though they are both in the 1-$\sigma$ of the estimated metallicity. However, the few dwarfs of NGC 6205 present in the SDSS/SEGUE catalogue, have metallicities between $-2<$[Fe/H]$<-1.5$ with a peak around 1.7. Therefore it is more likely that the under-estimation of the metallicity is related to the SDSS/SEGUE metallicity than directly related to our algorithm.

When we applied the algorithm to any stars in the field, it is not possible to remove the sub-giant contamination. The distances predicted by the algorithm for dwarfs in the four globular clusters without the constraint on the intrinsic absolute magnitude of the dwarf to be larger than 3.7 are shown in Table \ref{table_const}. Releasing this constraint reduces systematically the predicted distance of the globular clusters. However, these values are still consistent with the distance found in the literature, with the exception of NGC 5466. Due to its distance, the fraction of sub-giants/giants is larger than in the other globular clusters, leading to a more significant impact of these stars on the distance estimation. This bias should be considered when working with the dwarf stars at large distance (typically $>10$ kpc). Interestingly, taking into account the sub-giant contamination has a very little impact on the estimated metallicity, the difference with the metallicity found previously being much smaller than the scatter found with the spectroscopic measurement.

\begin{table}[ht]
 \centering
  \caption{Predicted mean distances and metallicity of the dwarfs in the 4 globular clusters without the constraint on the intrinsic absolute magnitude.}
  \label{table_const}
  \begin{tabular}{@{}l|cc@{}}
  \hline
   Name & $D_{dwarfs}$ (kpc) & [Fe/H]$_{dwarfs}$\\
    \hline
NGC6205 & $7.9 \pm{1.7}$ & $-1.69 \pm{0.3}$\\
NGC6341 & $8.5 \pm{2.1}$ & $-2.0 \pm{0.37}$\\
NGC5272 & $9.5 \pm{2.2}$ & $-1.77 \pm{0.36}$\\
NGC5466 & $11.9 \pm{3.9}$ & $-1.78 \pm{0.36}$\\
\hline
\end{tabular}
\end{table}

In addition to these relatively nearby clusters, NGC 2419 and the Draco dwarf spheroidal (dSph) are also present in the CFIS footprint. These systems are much more distant, at $79.7 \pm{0.3}$ kpc and $74.26 \pm{0.18}$ kpc\footnote{These uncertainties do not include the systematics.},  respectively \citep{hernitschek_2019}. As such, only the upper portion of the giant branch in these two satellites are visible in our data, and most of these stars have relatively poor photometric measurements. This leads to large uncertainties on the distances predicted by our algorithm. In what follows, we only use stars whose intrinsic absolute magnitudes are known to better than $\delta M_{G} \leq 0.5$. This corresponds to a precision on the distance of at least 23\%. The left panels of Figure \ref{draco_2419} show the CMDs for these two satellites, adopting the literature distance for the absolute magnitude on the y-axis. The middle panel only keeps stars which also have $\delta M_{G} \leq 0.5$.

We remind the reader that these two objects currently provide the best opportunity to test the validity of our technique at large distances, and were used to determine the global offset of the parallax for the giants, $\varpi_0 = 0.033$ mas, that we adopted in Section \ref{sec_giant}. Using this calibration, for NGC 2419, the mean distance of the giants according to our algorithm is $77.6 \pm{15.2}$ kpc, in agreement with the distance found by \citet{hernitschek_2019} of $79.7 \pm{0.3}$ kpc. For Draco, we find a mean distance of $73.4 \pm{9.9}$ kpc for the giants, consistent with the $74.26 \pm{0.18}$ kpc found by \citet{hernitschek_2019}. We note that a smaller offset of $\varpi_0 = 0.029$ mas (as used for the dwarfs, \citealt{lindegren_2018}) would lead to a distance for these two satellites closer to $\sim 50$ kpc.

Finally, the CFIS footprint contains a large portion of the GD-1 stream \citep{grillmair_2006a}. Applying a similar proper motion selection as \citet{price-whelan_2018}, we measure a heliocentric distance for the stream of $8.0 \pm{1.5}$ kpc and a mean metallicity of [Fe/H]$=-1.9 \pm{0.4}$ consistent with the recent measurement of \citet{malhan_2018a}.

\begin{figure}
\centering
  \includegraphics[angle=0, clip, width=8.0cm]{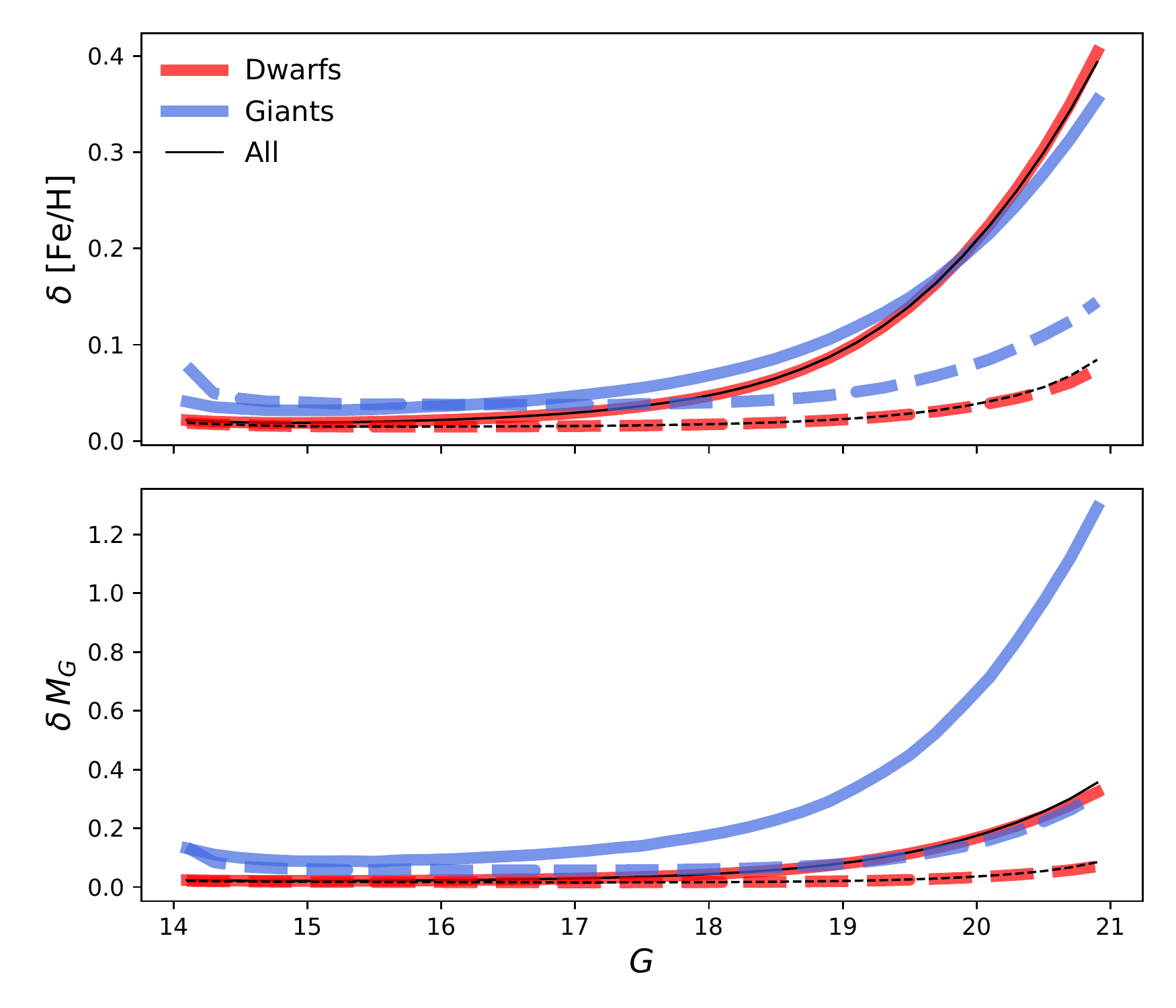}
   \caption{{\it Upper panel:} Distribution of the systematic error (continuous lines) and of the uncertainties due to the photometry (dashed lines) on the metallicity for the stars identified as dwarfs (in red), giants (in blue) and for the entire sample in black as function of the apparent magnitude in the $G$-band. {\it Lower panel:} similar to the upper panel for the absolute magnitude in the $G$-band.}
\label{error_distri}
\end{figure}

\subsection{Systematic and random uncertainties for the entire catalog} \label{all_cat}

In this section we apply the algorithm to the $\sim 12.8$ million stars present in the current version of the merged CFIS-PS1-Gaia catalog. The catalogue contains $12.2$ million stars identified as dwarfs ($P_{Dwarf}>0.5$) and $600,000$ stars ($\sim 5\%$) identified as giants ($P_{Giant}>0.5$). The distribution of the systematic errors ($\delta$[Fe/H]$_{sys}$ and $\delta M_{G,\,sys}$) and of the uncertainties due to the photometry (that we will call photometric uncertainties for simplicity; $\delta$[Fe/H]$_{photo}$ and $\delta M_{G,\,photo}$) on the metallicity and the absolute magnitude as a function of the apparent magnitude in the $G$-band are presented in Figure \ref{error_distri}. The systematic errors and photometric uncertainties for the dwarfs are lower than for the giants, and it is interesting to see that the systematic errors are generally higher than the photometric uncertainties, especially for stars fainter than $G=17.5$. This indicates that the relations found by each independent ANN (that give similar results for the training and testing sample) predict different values at fainter magnitudes. The differences between these values are more important than the uncertainties on these parameters due to photometric uncertainties. These differences are likely a consequence of the low number of stars in the training and testing samples below $G=18$ ($<10\%$), which has caused the ANNs to determine relationships based mainly on the  brighter stars. We expect that a training set based on the next generation of spectroscopic surveys, such as WEAVE, 4-MOST or SDSS-V, will be able to significantly reduce the systematic errors at faint magnitude, since these catalogues will contain a higher number of stars at fainter magnitudes than SDSS/Segue.

We recommend to only use stars classified as dwarfs or giants with high confidence (i.e. $P_{Dwarf}\geq 0.7$ or $P_{Giant}\geq 0.7$), and which have uncertainties on their estimated absolute magnitude $\delta {M_G} = \sqrt{(\delta M_{G,\,sys}^2+\delta M_{G,\,photo}^2)} \leq 0.5$. This selection remove most of the contamination from the mis-classified dwarfs/giants while keeping a large number of stars whose distances and metallicities are ``correct'' (with a relative uncertainty on the individual distances less than $26\%$, and on the individual metallicities of less than $0.3$ dex).

\section{The metallicity distribution of the outer Galaxy} \label{map_metal}

\begin{figure*}
\centering
  \includegraphics[angle=0, clip,  viewport= 0 0 755 770, width=17.5cm]{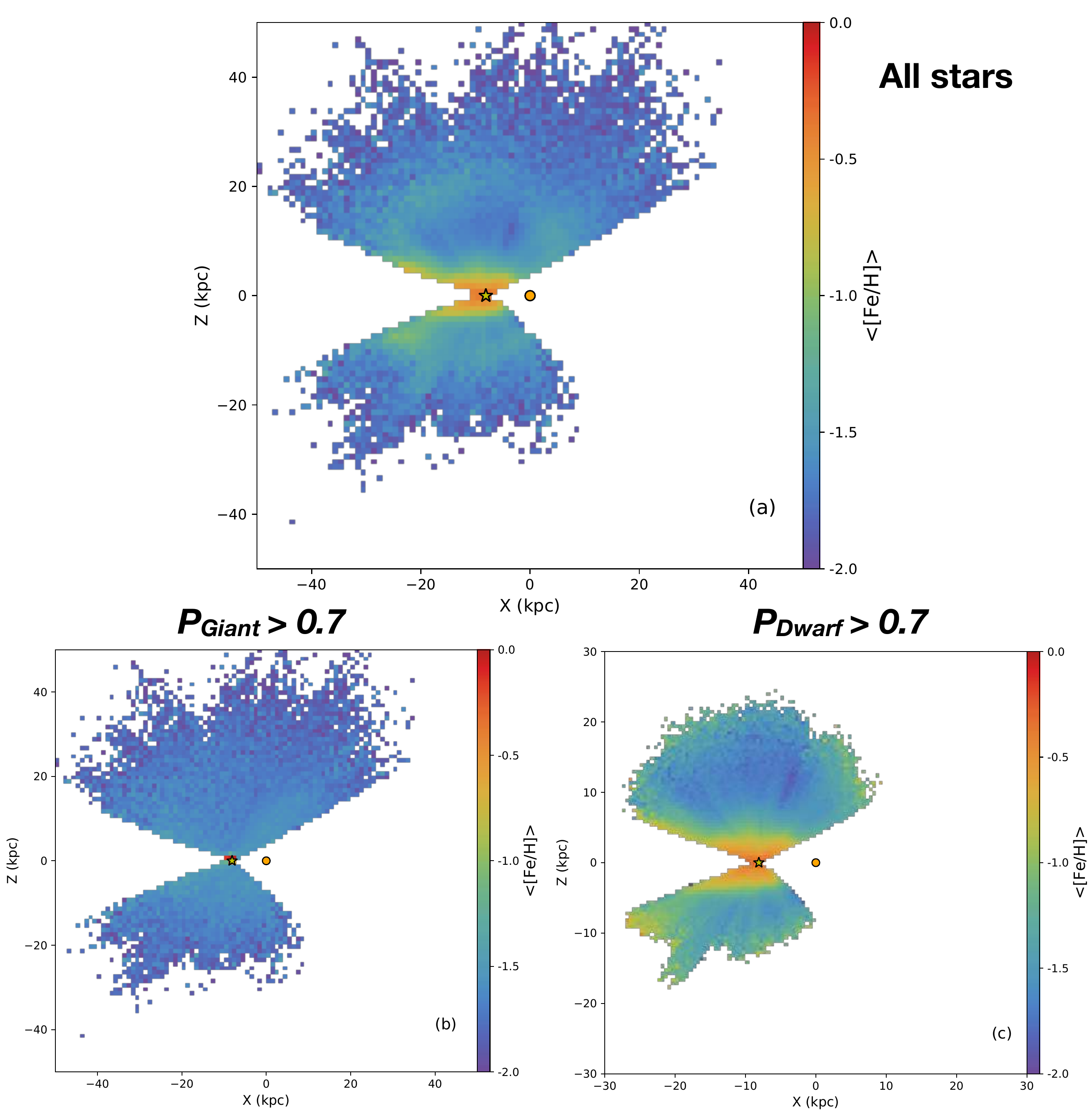}
   \caption{(a) The spatial variation of the mean metallicity of all the stars in the CFIS/PS1/Gaia dataset confidently classified by our algorithm as dwarf or giant (i.e., $P\ge 0.7$), a total of $\simeq 11.3$ million objects. The map is composed of 2,500 pixels ($1$ kpc $\times 1$ kpc), with a minimum of 5 stars per pixel. (b) Same as (a), but only for the giants, $ \simeq 136,000$ million stars (c) Same as (a), but only for the dwarfs, $\simeq 11.2$ million stars. This last map is composed of 3,600 pixels (0.5 kpc $\times$ 0.5 kpc), with still a minimum of 5 stars per pixel. In each panel, the yellow star shows the position of the Sun and the orange circle indicates the position of the Galactic center. No completeness corrections have been applied; thus, for the dwarfs, the increase of metalicity at large radius is an artefact of the fact that more metal rich dwarfs are brighter, and thus visible to larger radius, than metal poor dwarfs.}
\label{distri_stars}
\end{figure*}

The primary purpose of developing the algorithm described and tested here, is to be able to analyze the structure of the Galaxy using individual stellar metallicities and distances for the millions of stars present in photometric catalogs. Specifically, we have applied our algorithm to the $\sim 12.8$ million stars present in the current version of the merged CFIS-PS1-Gaia catalog. A detailed analysis of various aspects of this dataset as it relates to Galactic structure will be presented in future papers. Here, we provide a glimpse of the opportunities for studies of the outer Galactic science by showing projections of the spatial variation in the mean metallicity in Figure \ref{distri_stars}. We only use stars classified as dwarfs or giants with high confidence (i.e. $P_{Dwarf}\geq 0.7$ or $P_{Giant}\geq 0.7$), and which have uncertainties on their estimated absolute magnitude $\delta {M_G} \leq 0.5$, which are our recommended selection, as mentioned in Section \ref{all_cat}. 

The projections in Figure \ref{distri_stars} use the Galactic Cartesian coordinates ($X$, $Y$, $Z$), defined using the conventions adopted in the \textsf{astropy} package \citep{theastropycollaboration_2018}, with the most recent estimate of the Sun's position ($X = -8.1$ kpc, \citealt{gravitycollaboration_2018a}). {Pixel sizes are  $1$ kpc $\times 1$ kpc for panels $(a)$ and $(b)$ and of $0.5$ kpc $\times 0.5$ kpc for the panel $(c)$. Each pixel contain at least 5 stars to remove the impact of isolated stars.} For the dwarfs, we see that the mean metallicity apparently increases at $\sim 20$ kpc. This effect is not physical and is an artifact caused by the limited photometric depths of the different catalogs, especially Gaia and the $z$-band of PS1. At a given apparent magnitude, the metal-rich dwarfs are intrinsically brighter than more metal-poor dwarfs of the same age. Therefore metal-rich dwarfs are detected to larger distance than the metal poor-stars, leading to an artificial increase of the mean metallicity with distance. This effect explain the presence of the apparent metal-rich structure present in panel $(a)$ at $\sim 20$ kpc, since dwarfs are $\sim 3$ times more numerous than giants at this distance.

\section{Discussion and Conclusions} \label{conclusion}

In this paper, we present a new data-driven method based on Machine Learning algorithms that effectively distinguishes dwarfs from giants, and estimates distances and metallicities to each, using multi-wavelength photometry in the optical/near-infrared regime. It is based on CFIS, PS1, and Gaia photometry, but the general principals are applicable to any multi-band dataset.

Using this technique, we were able to recover more than 50\% of the giants observed in SDSS/SEGUE, with a contamination from misidentified dwarfs lower than 30\%. This technique works best for low metallicities, [Fe/H]$\leq  -1.2$, for which more than 70\% of the giants present in SDSS/SEGUE are correctly identified. Due to the low number of metal-poor dwarfs in our training set in the temperature range occupied by both dwarfs and giants, the predicted sample of giants may be contaminated by a non-negligible fraction of true dwarfs. However, using the Gaia proper motion, a majority of this contamination can be removed. 

{Hopefully, the new spectroscopic survey in the northern hemisphere, such as WEAVE or SDSS-V, will increase this number of metal-poor dwarfs and will provide an excellent training set to improve the dwarfs/giants classification, whose the principles have be developed in this paper.}

We obtain accurate photometric metallicities for both dwarfs and giants, with a scatter with respect to the spectroscopic measurements of 0.15 and 0.16 dex, respectively. We also obtain good estimates of the distances to the dwarfs and giants; we estimate a distance precision of the galactic objects of typically $\simeq 18$\% for dwarfs and $\simeq 26$\% for giants. In contrast to other techniques (e.g., \citet{bailer-jones_2015,bailer-jones_2018}), we do not require any prior on their distance distribution. This is critical, since it allows us to use the distances to analyze the spatial distribution of stars in the distant Galaxy, as well as their metallicities. Compared to previous data-driven methods \citep[e.g.][]{juric_2008,ivezic_2008,ibata_2017b}, the algorithm presented in this paper is able discriminate dwarfs and giants for stars with a metallicity lower than [Fe/H]$=-1.0$ and the distance callibration is done using the tremendous number of parallaxes obtain by the Gaia mission. Using our algorithm, we realistically account for the uncertainties due to photometric errors in addition to systematic errors.

As part of this analysis, we publish our training/test set for the publicly-available part of our photometric dataset (specifically, in the $2,608$ deg$^2$ obtained in the 2015 - 2017 observing period and which is colored orange in Figure~\ref{footprint}) and is available on the CADC web site: {\it www.canfar.net/storage/list/gthomas/dwarfOrGiants.fits}. This catalog contains all relevant photometric, spectroscopic, astrometric and derived parameters, as described in Table \ref{cat_online}.

The critical and major limitation of this method is attributable to the representative nature of our training/test sets, specifically SDSS/SEGUE, and Gaia. By construction, our algorithm assumes that the stars in our dataset represent the full diversity and totality of all stars in the Galaxy (especially at the high latitudes targeted by CFIS). This assumption necessarily introduces biases, which we quantify. Most notably, our algorithm assumes all stars are either dwarfs or red giant branch stars, and our techniques work best at [Fe/H]$ < -1.2$ dex. At higher metallicities, our training set has an absence of metal-rich giants. In addition, at very low metallicities, there is an deficit of metal-poor dwarfs in our training set that have temperatures in the same range as the giants. Stars with these specific metallicities (and temperatures), as well as other types of giants (such as the asymptotic giant branch stars) are not present in the SDSS/SEGUE dataset with sufficient frequency to allow our current techniques to overcome these limitations. However, we demonstrate with tests using LAMOST data and Galactic satellites that these biases are well understood and that the results of our algorithm can be used with confidence for statistical studies of distances and metallicities in the distant and metal-poor Galaxy. 

Future papers will analyse the spatial structure of the outer Galaxy and its metallicity distribution from the full CFIS-PS1-Gaia catalog. Crucially, the methodology that we use to connect SDSS/SEGUE spectroscopy to CFIS-PS1 photometry can be used in the future to connect imminent spectroscopic surveys such as WEAVE, 4MOST, SDSS-V and DESI to LSST and UNIONS, opening up an unprecedented discovery space for Milky Way stellar population studies. Critical to the success of these efforts will be ensuring well-defined spectroscopic training sets that sample a broad range of stellar parameters with minimal biases.

\section*{Acknowledgements}
NFM and RI gratefully acknowledge support from the French National Research Agency (ANR) funded project ``Pristine'' (ANR-18-CE31-0017) along with funding from CNRS/INSU through the Programme National Galaxies et Cosmologie and through the CNRS grant PICS07708. ES gratefully acknowledges funding by the Emmy Noether program from the Deutsche Forschungsgemeinschaft (DFG).

This work is based on data obtained as part of the Canada-France Imaging Survey (CFIS), a CFHT large program of the National Research Council of Canada and the French Centre National de la Recherche Scientifique. Based on observations obtained with MegaPrime/MegaCam, a joint project of CFHT and CEA Saclay, at the Canada-France-Hawaii Telescope (CFHT) which is operated by the National Research Council (NRC) of Canada, the Institut National des Science de l'Univers (INSU) of the Centre National de la Recherche Scientifique (CNRS) of France, and the University of Hawaii, and on data from the European Space Agency (ESA) mission {\it Gaia} (\url{https://www.cosmos.esa.int/gaia}), processed by the {\it Gaia} Data Processing and Analysis Consortium (DPAC,
\url{https://www.cosmos.esa.int/web/gaia/dpac/consortium}). Funding for the DPAC has been provided by national institutions, in particular the institutions participating in the {\it Gaia} Multilateral Agreement.

We also used the data provide by the Guoshoujing Telescope (the Large Sky Area Multi-Object Fiber Spectroscopic Telescope LAMOST), a National Major Scientific Project built by the Chinese Academy of Sciences. Funding for the project has been provided by the National Development and Reform Commission. LAMOST is operated and managed by the National Astronomical Observatories, Chinese Academy of Sciences.\\

Funding for the Sloan Digital Sky Survey IV has been provided by the Alfred P. Sloan Foundation, the U.S. Department of Energy Office of Science, and the Participating Institutions. SDSS-IV acknowledges support and resources from the Center for High-Performance Computing at the University of Utah. The SDSS website is http://www.sdss.org. SDSS-IV is managed by the Astrophysical Research Consortium for the Participating Institutions of the SDSS Collaboration including the Brazilian Participation Group, the Carnegie Institution for Science, Carnegie Mellon University, the Chilean Participation Group, the French Participation Group, Harvard-Smithsonian Center for Astrophysics, Instituto de Astrof\'{i}sica de Canarias, The Johns Hopkins University, Kavli Institute for the Physics and Mathematics of the Universe (IPMU)/University of Tokyo, Lawrence Berkeley National Laboratory, Leibniz Institut f\"{u}r Astrophysik Potsdam (AIP), Max-Planck-Institut f\"{u}r Astronomie (MPIA Heidelberg), Max-Planck-Institut f\"{u}r Astrophysik (MPA Garching), Max-Planck-Institut f\"{u}r Extraterrestrische Physik (MPE), National Astronomical Observatories of China, New Mexico State University, New York University, University of Notre Dame, Observat\'{a}rio Nacional/MCTI, The Ohio State University, Pennsylvania State University, Shanghai Astronomical Observatory, United Kingdom Participation Group, Universidad Nacional Aut\'onoma de M\'exico, University of Arizona, University of Colorado Boulder, University of Oxford, University of Portsmouth, University of Utah, University of Virginia, University of Washington, University of Wisconsin, Vanderbilt University, and Yale University.

The Pan-STARRS1 Surveys (PS1) have been made possible through contributions of the Institute for Astronomy, the University of Hawaii, the Pan-STARRS Project Office, the Max-Planck Society and its participating institutes, the Max Planck Institute for Astronomy, Heidelberg and the Max Planck Institute for Extraterrestrial Physics, Garching, The Johns Hopkins University, Durham University, the University of Edinburgh, Queen's University Belfast, the Harvard-Smithsonian Center for Astrophysics, the Las Cumbres Observatory Global Telescope Network Incorporated, the National Central University of Taiwan, the Space Telescope Science Institute, the National Aeronautics and Space Administration under Grant No. NNX08AR22G issued through the Planetary Science Division of the NASA Science Mission Directorate, the National Science Foundation under Grant No. AST-1238877, the University of Maryland, and Eotvos Lorand University (ELTE).

\bibliography{./biblio}

\appendix
\section{Description of the online catalogue}

\begin{table*}[h]
 \centering
  \caption{Description of each column present in the online catalogue}
   \label{cat_online}
  \begin{tabular}{@{}l|l|l@{}}
  \hline
   No & Column name & Description \\
    \hline
1 & RA & Right Ascension (deg) \\
2 & Dec & Declination (deg)\\
3 & u\_cfis & $u$-band photometry \\
4 & u0\_cfis & Deredded $u$-band photometry \\
5 & du\_cfis & Uncertainty on the $u$-band photometry \\
6 & g\_PS & PS1 mean $g$-PSF photometry \\
7 & g0\_PS & Deredded PS1 mean $g$-PSF photometry \\
8 & dg\_PS & Uncertainty on the $g$-band photometry \\
9 & r\_PS & PS1 mean $r$-PSF photometry \\
10 & r0\_PS & Deredded PS1 mean $r$-PSF photometry \\
11 & dr\_PS & Uncertainty on the $r$-band photometry \\
12 & i\_PS & PS1 mean $i$-PSF photometry \\
13 & i0\_PS & Deredded PS1 mean $i$-PSF photometry \\
14 & di\_PS & Uncertainty on the $i$-band photometry \\
15 & z\_PS & PS1 mean $z$-PSF photometry \\
16 & z0\_PS & Deredded PS1 mean $z$-PSF photometry \\
17 & dz\_PS & Uncertainty on the $z$-band photometry \\
18 & y\_PS & PS1 mean $y$-PSF photometry \\
19 & y0\_PS & Deredded PS1 mean $y$-PSF photometry \\
20 & dy\_PS & Uncertainty on the $y$-band photometry \\
21 & G\_gaia & Gaia $G$-band photometry \\
22 & G0\_gaia & Deredded Gaia $G$-band photometry \\
23 & dG\_gaia & Uncertainty on the Gaia $G$-band photometry \\
24 & BP\_gaia & Gaia $G_{BP}$-band photometry \\
25 & BP0\_gaia & Deredded Gaia $G_{BP}$-band photometry \\
26 & dBP\_gaia & Uncertainty on the Gaia $G_{BP}$-band photometry \\
27 & RP\_gaia & Gaia $G_{RP}$-band photometry \\
28 & RP0\_gaia & Deredded Gaia $G_{RP}$-band photometry \\
29 & dRP\_gaia & Uncertainty on the Gaia $G_{RP}$-band photometry \\ 
30 & parallax & Gaia parallax (mas) \\
31 & dparallax & Uncertainty on the Gaia parallax (mas)\\
32 & pmra & Gaia proper motion in right ascension direction  (mas/yr)\\
33 & dpmra & Uncertainty on the Gaia proper motion in right ascension direction (mas/yr)\\
34 & pmdec & Gaia proper motion in declination direction  (mas/yr)\\
35 & dpmdec & Uncertainty on the Gaia proper motion in declination direction (mas/yr)\\
36 & EBV & Extinction from \citep{schlegel_1998} \\
\hline
\end{tabular}
\end{table*}

\begin{table*}
 \centering
  \caption{Suite of Table \ref{cat_online}}
  \begin{tabular}{@{}l|l|l@{}}
  \hline
   No & Column name & Description \\
    \hline
37 & proba\_dwarf & Probability of being a dwarf ($P_{Dwarf}$) \\
38 & proba\_giant & Probability of being a giant ($P_{Giant}$) \\
39 & dproba & Uncertainty on the Probability of being a dwarf/giant ($\delta P$)\\
40 & feh\_pred & Predicted photometric metallicity (equal to feh\_dwarf if $P_{Dwarf}>0.5$ and to feh\_giant if $P_{Giant}>0.5$)\\
41 & dfeh\_pred\_sys & Systematic error on the predicted photometric metallicity\\
42 & dfeh\_pred\_photo & Photometric uncertainty on the predicted photometric metallicity\\
43 & MG\_pred & Predicted absolute magnitude (equal to MG\_dwarf if $P_{Dwarf}>0.5$ and to MG\_giant if $P_{Giant}>0.5$)\\
44 & dMG\_pred\_sys & Systematic error on the predicted absolute magnitude\\
45 & dMG\_pred\_photo & Uncertainty on the predicted absolute magnitude\\
46 & feh\_dwarf & Predicted photometric metallicity using the dwarf calibration (Section \ref{sec_dwarf})\\
47 & dfeh\_dwarf\_sys & Systematic error on the predicted photometric metallicity of the dwarfs\\
48 & dfeh\_dwarf\_photo & Photometric uncertainty on the predicted photometric metallicity of the dwarfs\\
49 & MG\_dwarf & Predicted absolute magnitude using the dwarf calibration (Section \ref{sec_dwarf})\\
50 & dMG\_dwarf\_sys & Systematic error on the predicted absolute magnitude of the dwarfs\\
51 & dMG\_dwarf\_photo & Uncertainty on the predicted absolute magnitude of dwarfs\\
52 & feh\_giant & Predicted photometric metallicity using the giant calibration (Section \ref{sec_giant})\\
53 & dfeh\_giant\_sys & Systematic error on the predicted photometric metallicity of the giants\\
54 & dfeh\_giant\_photo & Photometric uncertainty on the predicted photometric metallicity of the giants\\
55 & MG\_giant & Predicted absolute magnitude using the giant calibration (Section \ref{sec_giant})\\
56 & dMG\_giant\_sys & Systematic error on the predicted absolute magnitude of the giants\\
57 & dMG\_giant\_photo & Uncertainty on the predicted absolute magnitude of giants\\
58 & dist\_pred & Heliocentric distance using the predicted absolute magnitude (kpc)\\
59 & MG\_gaia & Absolute magnitude computed by Equation \ref{eq_MG} from the Gaia parallaxes\\
60 & dMG\_gaia & Uncertainty on the absolute magnitude computed by Equation \ref{eq_dMG} from the Gaia parallaxes\\
61 & objid & Object ID from SDSS/SEGUE\\
62 & specobjID & Spectrocopic ID from SDSS/SEGUE\\
63 & Teffadop & Adopted effective temperature from SDSS/SEGUE (in Kelvin)\\
64 & Teffadopunc & Uncertainty on the adopted effective temperature from SDSS/SEGUE (in Kelvin)\\
65 & loggadop & Adopted surface gravity from SDSS/SEGUE\\
66 & loggadopunc & Uncertainty on the adopted surface gravity from SDSS/SEGUE\\
67 & FeHadop & Adopted spectroscopic metallicity from SDSS/SEGUE\\
68 & FeHadopunc & Uncertainty on the adopted spectroscopic metallicity from SDSS/SEGUE\\
69 & snr & Spectroscopic signal-to-noise ratio from SDSS/SEGUE\\
\hline
\end{tabular}
\end{table*}

\end{document}